%% file: 0749MAIN.tex
\documentclass{aa}             

\input epsf

\begin{document}

\def\kms{\ifmmode{\rm km\thinspace s^{-1}}\else km\thinspace s$^{-1}$\fi}
\def\feh{$[\mathrm{Fe/H}]$}
\def\meh{$[\mathrm{M/H}]$}
\def\afe{$[\mathrm{\alpha/Fe}]$}

\title{
A new standard: 
Age and distance for the open cluster \\
NGC\,6791 from the eclipsing binary member V20
\thanks{Based on observations carried out at 
Nordic Optical Telescope at La Palma and ESO's VLT/UVES
ESO, Paranal, Chile (75.D-0206A, 77.D-0827A).}
\thanks{Tables 11 and 12 are only available in electronic form
at the CDS via anonymous ftp to cdsarc.u-strasbg.fr
(130.79.128.5) or via http://cdsweb.u-strasbg.fr/cgi-bin/qcat?J/A+A/}
} 
\author{
F. Grundahl       \inst{1} 
\and J.V. Clausen \inst{2}
\and S. Hardis    \inst{2}
\and S. Frandsen  \inst{1}
}
\offprints{F. Grundahl, \\ e-mail: fgj@phys.au.dk}

\institute{
Department of Physics and Astronomy,
Aarhus University, 
Ny Munkegade, DK-8000 Aarhus C, Denmark
\and
Niels Bohr Institute, University of Copenhagen,
Juliane Maries Vej 30,
DK-2100 Copenhagen {\O}, Denmark
}

\date{Received / Accepted } 
 
\titlerunning{V20 in the open cluster NGC\,6791}
\authorrunning{F. Grundahl et al.}

\abstract
{We wish to determine accurate ages for open clusters and use this, 
 in conjunction with colour-magnitude diagrams, to constrain models 
 of stellar structure and evolution.}
{The detached eclipsing binary V20 in the old, metal--rich 
 (\feh\,$=+0.40$) open cluster NGC\,6791 is studied in order to determine 
 highly accurate masses and radii of its components. This allows the 
 cluster age to be established with high precision, using isochrones in 
 the mass-radius diagram.}
{We employ high-resolution UVES spectroscopy of V20 to determine the 
  spectroscopic orbit and time-series $V, I$ photometry to obtain the 
  photometric elements.}
{The masses and radii of the V20 components are found to be 
 $1.074\pm0.008M_{\sun}$ and $1.399\pm0.016R_{\sun}$ (primary)  and 
 $0.827\pm0.004M_{\sun}$ and $0.768\pm0.006R_{\sun}$ (secondary). 
 The primary is located almost exactly at the hottest point along the cluster 
 isochrone, and the secondary is a $\sim7$ times fainter main--sequence star.
 We determine an apparent cluster distance-modulus of 
 $(m-M)_V\,=\,13.46\pm0.10$ (average of primary and secondary). The 
 cluster age is obtained from 
 comparisons with theoretical isochrones in the mass--radius diagram. 
 Using the isochrones from Victoria--Regina with \feh\,$=+0.37$ we 
 find $7.7\pm0.5$Gyr, whereas the Yonsei-Yale ($Y^2$) isochrones lead to 
 $8.2\pm0.5$Gyr, and BaSTI isochrones to $9.0\pm0.5$Gyr.
 In a mass-radius diagram, the 7.7Gyr VRSS and 9.0Gyr BaSTI isochrones 
 overlap nearly perfectly despite the age-difference. This 
  model dependence, which is significantly larger than the precision
 determined from mass, radius, and abundance uncertainties, prevents a 
 definitive age-determination of the cluster.}
{Using detached eclipsing binaries for determination of cluster ages, the
 dominant error is due to differences among stellar models and no longer
 to observational errors in cluster reddening and distance. By observing 
 a suitable number of detached eclipsing binaries in several open clusters 
 it should be possible to calibrate the age--scale and provide firm 
 constraints which stellar models must reproduce.}
\maketitle

\section{Introduction}
\label{sec:intro}

 Among the old open clusters, NGC\,6791 holds a special place: it is one of  
 the oldest, most massive, and most metal-rich (Origlia \cite{origlia06}, 
 Carretta, Bragaglia \& Gratton \cite{carretta07}, Anthony--Twarog, Twarog 
 \& Meyer \cite{attm07}, hereafter ATTM07) clusters known. 
 In addition to these features, the 
 cluster contains a population of hot blue stars (Liebert, Saffer \& Green 
 \cite{liebert94}, Landsman et al. \cite{landsman98} ), a large population 
 of white dwarfs extending to the end of the cooling sequence  
 has been found with HST (Bedin et al. \cite{bedin05},  Bedin et al. 
 \cite{bedin08-1}), and extensive photometric studies have revealed a large 
 population of variable stars in the cluster field (Bruntt et al. \cite{hb03},
 Mochejska et al. \cite{m02}, De Marchi et al.  \cite{demarchi07}).

 For these reasons, NGC\,6791 has been the subject of a number of studies 
 since the work by Kinman (\cite{kinman65}), and yet there is still not
 agreement over its basic parameters. Given a distance of $\sim4\,$kpc 
 and a non-negligible reddening, determination of the cluster age is 
 a complex problem when attempting to use the colour-magnitude diagram  
 (hereafter CMD).  Even with recent attempts to detect exo-planet transits 
 (Bruntt et al. \cite{hb03}, Mochejska et al. \cite{mochejska05},
 Montalto et al. \cite{montalto07}) -- which has led to a substantial 
 body of well calibrated photometry and a high-precision CMD (Stetson, 
 Bruntt \& Grundahl \cite{stetson03}, hereafter SBG03) -- 
 it has proven very difficult to determine a precise cluster age. 
 This is mainly related to the difficulty of obtaining precise reddening 
 and distance estimates (see ATTM07 for an extensive discussion) and to the 
 problem of transforming model isochrones to observed colours and magnitudes 
 at the high metallicity of the cluster (Tripicco et al. \cite{tripicco95}). 

 The range of ages proposed for NGC\,6791 has extremes of 7 and 12Gyr,
 but has in recent years seemed to converge on a value near 8Gyr. This 
 would put the cluster turnoff mass close to 1$M_{\sun}$ and thus bridge the
 ``gap'' between the turnoff mass of globular clusters (typically
 0.8$M_{\sun}$) and younger open clusters, where convective overshoot
 has a marked effect on their turnoff morphology in the CMD. Such a high
 cluster age would also seem to suggest that atomic diffusion might have
had sufficient time to act in its stars, such as found in the globular
cluster NGC~6397 (Korn et al. \cite{korn06}). 

 In the course of a study to try and detect exoplanetary transits in the 
 cluster (Bruntt et al. \cite{hb03}), we realized that the detached 
 eclipsing system V20 (Rucinski, Kaluzny \& Hilditch \cite{rucinski96}) 
 might offer the possibility to determine
 the masses and radii for stars near the turnoff. Rough estimates of the 
 system parameters (Bruntt et al. \cite{hb03}) suggested that although V20 is
 only at $V=17.34$, it would still be within the capabilities of UVES 
 (Dekker et al. \cite{dekker00}) at the ESO VLT to measure precise radial 
 velocites for its components.

 It is well known that detached eclipsing binaries 
 offer the possibility to determine precise masses and radii for the 
 system components. If one or both components is close to its 
 turnoff mass (for the age of the binary), it is possible to put tight
 constraints on the age of the system through a comparison  of the 
 position of the primary and secondary in a mass--radius ($M-R$) diagram to 
 theoretical isochrones. For stellar clusters, such an analysis has some
 significant advantages: the determination of the masses and radii is 
 essentially independent of the usual ``trouble-makers''  
 such as reddening, distance, metallicity. Furthermore, since the 
 comparison to models is carried out in the $M-R$ diagram, one avoids
 the difficult process of transforming the effective temperatures and 
 luminosities of the models to observed colours and magnitudes. Thus, a
 determination of cluster ages in the $M-R$ diagram is essentially
 ``the closest'' one can get to a ``direct'' confrontation between 
 observations and models. 

 In this paper we undertake a full analysis of the detached eclipsing 
 binary system V20 in NGC\,6791 and determine accurate values for the 
 masses and radii of its components.
 This is used to determine a precise cluster age and to compare the
 ages derived using commonly used isochrones from three sources. 
 Our 
 main conclusion is that using such binary systems as V20, it is 
 the ``accuracy'' of the available theoretical stellar models, which 
 limit the obtainable age precision, and not the observational data. 

\section{Photometry of V20}
\label{sec:phot}

The photometric data for \object{V20} consists of $V$ (Johnson) 
and $I$ (Cousins) CCD observations from the 2.56m Nordic Optical 
Telescope (NOT) and its stand-by camera StanCam. We refer to the
telescope homepage for further 
information\footnote{{\scriptsize\tt http://www.not.iac.es}}. 
V20 was observed on 9 nights between April 2003 and July 2004,
and a total of 298 and 300 exposures were obtained in $V$ and $I$,
respectively.
Primary eclipse observations were done on three nights, secondary eclipse
observations on four nights, and out-of-eclipse phases on
two nights. Although complete coverage of the light curve has not been
secured outside eclipses, the data are sufficient for a 
detailed photometric analysis.

For all observations we employed an exposure time of 240s in $V$ and 
180s in $I$. Seeing conditions varied substantially from night to night, 
with median values of 1\farcs00 and 0\farcs82 in $V$ and $I$, respectively. 
The best seeing frames have a $FWHM$ of 0\farcs60 and 0\farcs46,
respectively. 

The bias frames and flat fields, used in the data reduction, were obtained 
during evening twilight on each observing night.  All photometry was 
carried out with DAOPHOT/ALLSTAR/ALLFRAME (Stetson \cite{stetson87, 
stetson94}) and DAOGROW (Stetson \cite{stetson90}) and transformed 
to a common coordinate system using MATCH and MASTER 
(P. Stetson, private comm.). For each frame we
produced a point-spread function (PSF) using the brightest stars in the
field, and subsequently carried out aperture photometry in large apertures
(neighbour stars subtracted) using the NEDA routine provided with DAOPHOT.
Subsequently the aperture photometry was fed into DAOGROW to obtain the
final large-aperture magnitudes. We found that this prodedure gives slightly
better photometric precision than using profile-fitting photometry.

\begin{figure}
\epsfxsize=90mm
\epsfbox{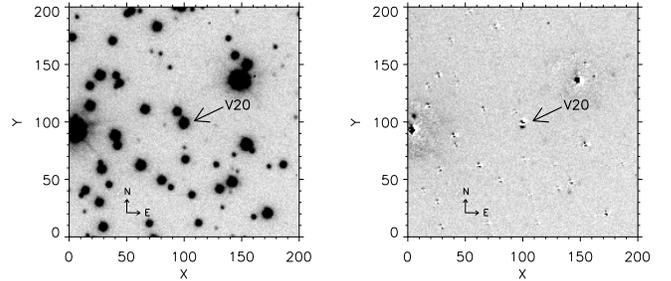}
\caption[]
{
\label{fig:v20_image}
The field of V20 in NGC 6791, centred at 
$(\alpha,\delta)_{2000}\,=\,(19^h\,20^m\,54.30^s +37^d\,45^m\,34.7^s)$.
Note, that in the left image V20 appears slightly elongated in 
the N-S direction, the FWHM in this $I$ band image is 0\farcs46. 
The righthand image shows the star subtracted version of the 
left image. Note the larger than expected residuals at the position
of V20, consistent with a fainter third light component. The pixel 
size is 0\farcs1755 and the greyscale is identical for the two 
images. 
}
\end{figure}

During the reductions of the data, we noticed on the best seeing frames
($FWHM\,=\,0\farcs46)$ that V20 appeared slightly elongated, and we interpret
this as a signature of a third light component due to a chance alignment with
a cluster star. This has to be properly taken into account in the final 
analysis of the lightcurves. In Fig.\ref{fig:v20_image} we show the field in 
NGC\,6791 centred on V20. Attempts were made to carry out ALLFRAME photometry
with two components at the position of V20 separated by 0\farcs4. This 
improved the residuals in the fit for the best seeing frames, but
the large seeing variations did not allow a sufficient precision in the 
magnitudes of the V20 primary$+$secondary and the third light component. 
We therefore include the light from the third star in the light
curves, and deal with it in the subsequent analyses.

\subsection{Light curves and standard indices}
\label{sec:lc}

With the instrumental photometry in hand, we proceeded to transform
the observations to the $V$ and $I$ standard system. In 
SBG03 a large effort was put into 
this transformation, and we have used the available photometry from this
source as internal standard stars. For each frame, a linear transformation
from instrumental magnitudes to standard magnitudes, using $(V-I)$ as 
colour term, was calculated, and subsequently we averaged the coefficient
for the colour term for all frames (for a given filter) and used this 
for the final determination of the zeropoint for each frame. 
Our photometry is therefore on the same system as SBG03 
with the out-of-eclipse magnitudes given by their photometry as 
listed in Table~\ref{tab:v20_VI}.
The accuracy of the photometry is in the range 0\fm01 to 0\fm02 as 
mentioned in SBG03.

\begin{figure}
\epsfxsize=90mm
\epsfbox{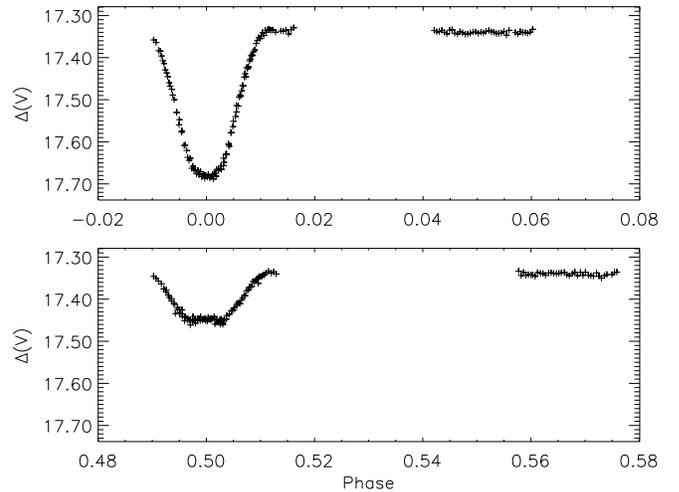}
\caption[]{\label{fig:v20_V}
$V$ light curve for V20 (third star included). 
Only the observed phase intervals are shown.
}
\end{figure}

\begin{figure}
\epsfxsize=90mm
\epsfbox{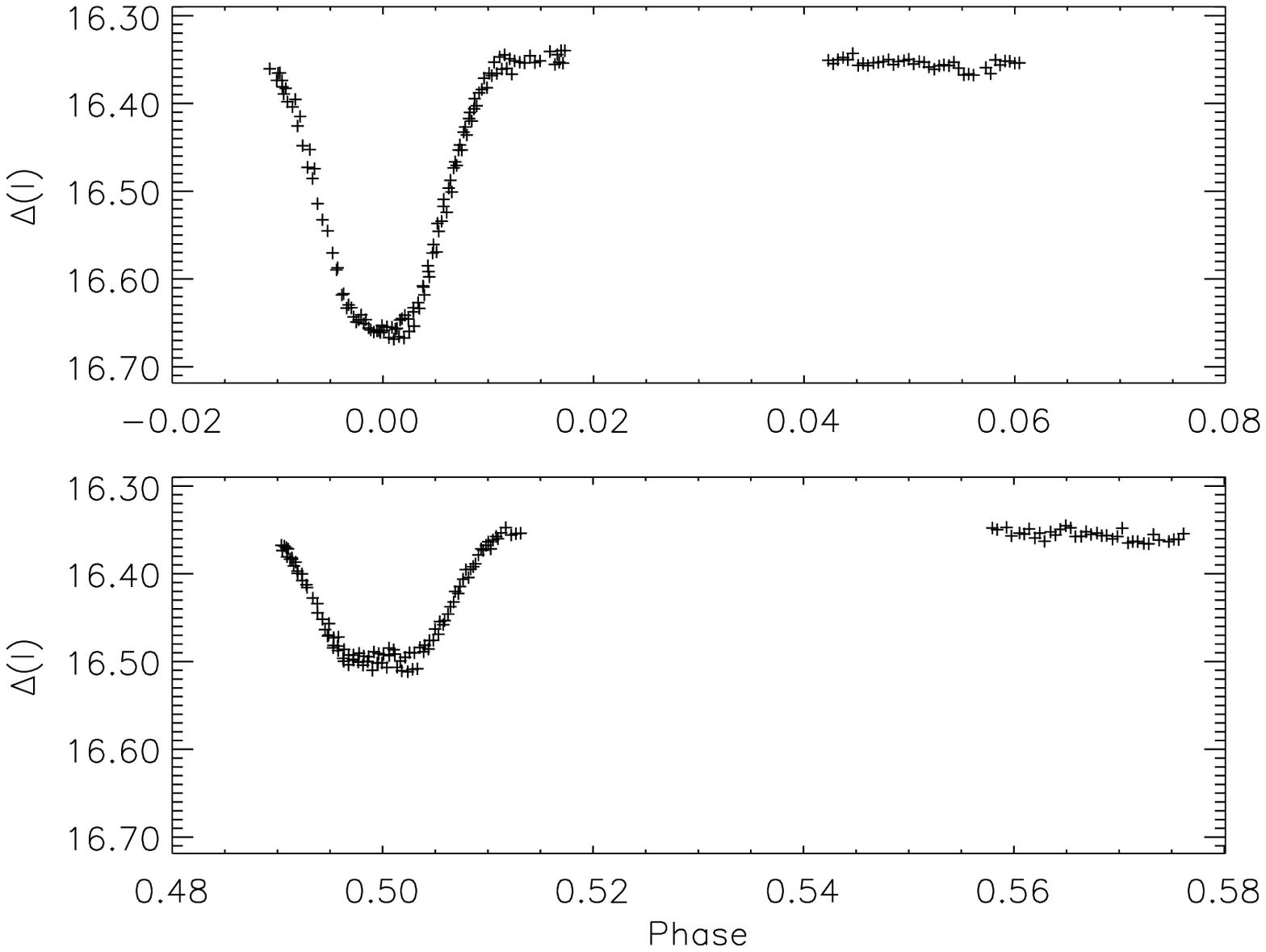}
\caption[]{\label{fig:v20_I}
$I$ light curve for V20 (third star included). 
Only the observed phase intervals are shown.
}
\end{figure}

The light curves for V20 are listed in Tables 11 and 12 and are shown in 
Figs.~\ref{fig:v20_V} and \ref{fig:v20_I}
with phases calculated from the ephemeris given in Eq.~\ref{eq:v20_eph}.
V20 is clearly well detached with practically constant light level outside
eclipses. Secondary eclipse occurs at phase 0.50, and the eclipses are of 
the same duration, supporting that the orbit is circular. 
As seen from the different eclipse depths, V20 consists of two stars of rather 
different surface fluxes.
Secondary eclipse is total; depths are $0\fm111$ ($V$) and $0\fm142$ ($I$).
The close companion is included in the light curves, meaning that a significant 
amount of third light is present. 

Throughout the paper, the component eclipsed at the deeper eclipse at 
phase 0.0 is referred to as the primary $(p)$, and the other as the 
secondary $(s)$ component.

Standard, out--of--eclipse, $V,I$ photometry for V20 is listed in 
Table~\ref{tab:v20_VI} together with calculated individual 
photometry for the three stars. In order to obtain constraints on 
the contribution from the third light, we shall attempt to 
determine its value from the available light curve. 

We first determined the depth of the total secondary eclipse in $V$ and $I$ 
to be 0\fm111 and 0\fm142 , respectively. Since this represents the case 
where the 
secondary component of the binary is totally eclipsed, we can determine 
the magnitudes for the secondary component, and for the sum of the primary 
and third light component. 
If we assume (as is verified later from spectroscopy) that V20 is a member 
of the cluster, and that the third light is also a cluster member, then by 
assuming that it is located on the main sequence, a unique combination can 
be calculated, for which the sum of primary and third light matches the light 
during total eclipse. 
In practical terms, we fitted a low order polynomial to the main-sequence of 
the cluster (representing possible locations for the third light component). 
Then, for each point along the polynomial, the corresponding location of the 
primary component was calculated, such that the sum of their light matched 
the location of the light during total eclipse of the secondary component. 

\begin{figure}
\epsfxsize=90mm
\epsfbox{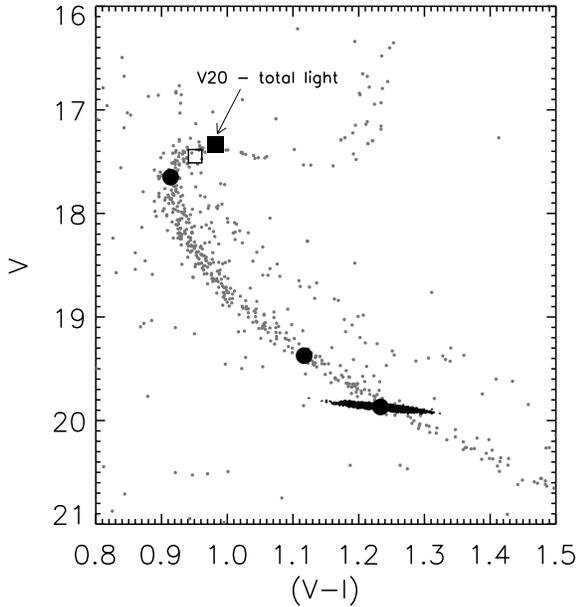}
\caption[]{\label{fig:v20_solved}
The $(V-I,V)$ colour-magnitude diagram for NGC\,6791 with V20 and its 
components overplotted. The filled square indicates the location of 
the total light of the entire system. The light of the primary and 
third component (light during total eclipse of the secondary) is 
indicated by an open square. In order of decreasing luminosity, the 
filled circles indicate the location of the primary, the third star,
and the secondary. For the secondary, we have overplotted the possible 
locations if errors of 0\fm002 in $V$ and $I$-depths of the total eclipse 
are assumed. 
}
\end{figure}

The results of this exercise is given in Table\ref{tab:v20_VI} for the primary, 
secondary and third star. In Fig.\ref{fig:v20_solved} we show the cluster 
$(V-I),V$ colour-magnitude diagram based on the Stetson et al. (2003) photometry 
with the location of the V20 combined light and the individual components 
indicated. We have attempted to estimate the uncertainty in the location of
the secondary component by assuming that the uncertainty in the estimate of the
$V$ and $I$ depths of the total eclipse is 0\fm002 and then calculating its 
position 10000 times at random from a normal distribution of $\sigma\,=0\fm002$ 
in each colour and calculating the position of the secondary. 
As can be seen from Fig.\ref{fig:v20_solved} it appears that three cluster
stars can indeed match the observed total light (as expected) and that the 
third light component has a luminosity in between the primary and secondary.

The exact luminosity for the third star is somewhat dependent on how the 
fiducial for the main-sequence is derived. We estimate that the uncertainty in
its determined position is not larger than 0\fm15 -- this is based on 
experiments where small shifts in colour were added to the fitted 
main-sequence polynomial, while it still appeared to represent the 
main-sequence. 
We note, that high-resolution imaging would allow a direct determination 
of the contribution to the total light from the third star. 
For the expected separation ($\sim$0\farcs4), this would probably 
require adaptive optics.


\input 0749tab1.tex

\subsection{Times of minima and ephemeris}
\label{sec:eph}

\input 0749tab2.tex

From the $V$ and $I$ light curve observations, two times of minima per band
have been derived for both eclipses.
They are listed in Table~\ref{tab:v20_tmin} together with times redetermined
from the photometry by Bruntt et al. (\cite{hb03}) and Mochejska et al. 
(\cite{m02}).
The method of Kwee and van Woerden (\cite{kvw56}) was applied throughout.

Weighted linear least square fits to the times of minima yield formally
identical periods of $14\fd46989 \pm 0.00005$ and $14.46999 \pm 0.00017$ 
for primary and secondary eclipses, respectively.
Since only three times per eclipse are available, we have independently
determined both the epoch and the period from JKTEBOP analyses of our
$V$ and $I$ light curves; see Sect.~\ref{sec:phel}. 
The periods obtained from the two light curves agree well and are close 
to those given above.
We adopt the following linear ephemeris for all analyses in this paper: 

\begin{equation}
\label{eq:v20_eph}
\begin{tabular}{r r c r r}
{\rm Min \, I} =  & 2453151.6061  & + & $14\fd 469918$ &$\times\; E$ \\
                  &       $\pm 9$ &   &        $\pm25$ &             \\
\end{tabular}
\end{equation}

\subsection{Photometric elements}
\label{sec:phel}

Since V20 is well detached with relative component radii of only
about 0.045 and 0.025, respectively, we have adopted the simple 
Nelson-Davis-Etzel model (Nelson \& Davis \cite{nd72}, 
Etzel \cite{e81}, Martynov \cite{m73}) for the light curve analyses. 
It represents the deformed stars as biaxial ellipsoids and applies a simple 
bolometric reflection model.
We have used the corresponding 
JKTEBOP\footnote{{\scriptsize\tt http://www.astro.keele.ac.uk/$\sim$jkt/}} code, which is a
revised and extended version of the original EBOP code (Etzel \cite{pe81}). 
The Levenberg-Marquardt minimization algorithm (MRQMIN: 
Press et al. \cite{press92})
is used for the lest-squares optimization of the parameters, and the code has
been extended to include also non-linear limb darkening and adjustment of
epoch and orbital period. In one of its modes, JKTEBOP performs Monte
Carlo simulations, which we use to assign realistic errors to the photometric
elements. For further information, we refer to e.g. Southworth et al. 
(\cite{sms04,szm04,betaaur}) and Bruntt et al. (\cite{psicen}).

The $V$ and $I$ light curves were analysed independently, with
equal weights assigned to all observations.
The magnitude at quadrature was always included
as an adjustable parameter, and the phase of primary eclipse was allowed to
shift from 0.0. In initial JKTEBOP analyses, the epoch and orbital period
was included as free parameters and then fixed at the values given
in Eq.~\ref{eq:v20_eph}; see Sect.~\ref{sec:eph}. A circular orbit
was assumed throughout, and
the mass ratio between the components was kept at the
spectroscopic value (Table~\ref{tab:v20_orbit}). 
Gravity darkening coefficients corresponding to
convective atmospheres were applied, and the simple bolometric reflection
model built into EBOP/JKTEBOP was used. Linear limb darkening coefficients were
either assigned from theoretical calculations (Van Hamme \cite{vh93}; 
Claret \cite{c00}) according to the effective temperatures, surface 
gravities, and abundance, or left free. 

In text and tables on photometric solutions we use the following symbols:
$i$ orbital inclination;
$r$ relative radius;
$k =  r_s/r_p$;
$u$ linear limb darkening coefficient;
$y$ gravity darkening coefficient;
$J$ {\it central} surface brightness {ratio};
$L$ luminosity;
$l_3$ third light fraction.

As mentioned in Sect.~\ref{sec:phot} and \ref{sec:lc}, 
a close star is included in the
light curve data. The corresponding amount of third light
$l_3 = L_3/(L_p+L_s+L_3)$, as calculated from
the information in Table~\ref{tab:v20_VI}, is $0.146 \pm 0.021$ ($V$)
and $0.168 \pm 0.024$ ($I$). Uncertainties are based on realistic
estimates of the accuracy of the magnitudes of the total light 
($\pm 0\fm010$), the depths of secondary eclipse ($\pm 0\fm002$),
and the magnitudes calculated for the companion ($\pm 0\fm15$). 
{ 
We have adopted these $l3$ results, which are based on the
realistic assumption that the third star is a main sequence cluster member,
since it turns out that the amount of third light is not well constrained
by the light curves; see below.
}

\begin{figure}
\epsfxsize=90mm
\epsfbox{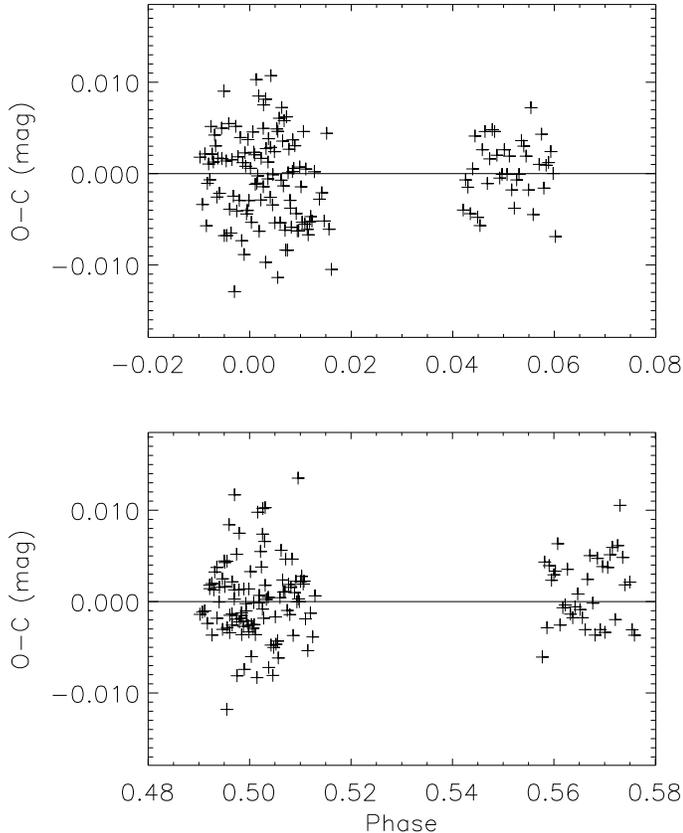}
\caption[]{\label{fig:v20_oc_v}
(O-C) residuals of the $V$ observations from the theoretical
light curves for the VH93 parameters listed in Table~\ref{tab:v20_ebop}.
}
\end{figure}

\begin{figure}
\epsfxsize=90mm
\epsfbox{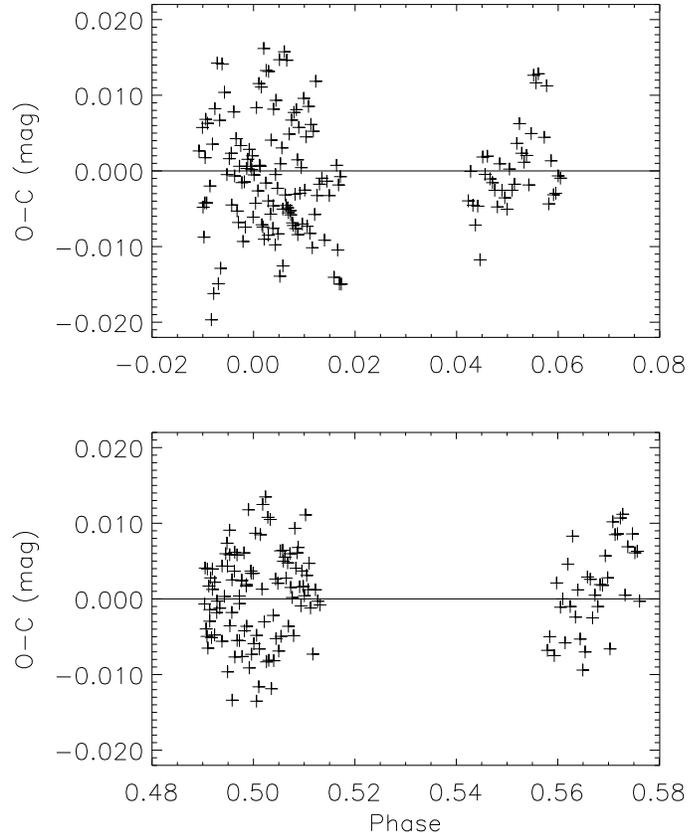}
\caption[]{\label{fig:v20_oc_i}
(O-C) residuals of the $I$ observations from the theoretical
light curves for the VH93 parameters listed in Table~\ref{tab:v20_ebop}.
}
\end{figure}

The photometric solutions for different adopted theoretical linear limb 
darkening coefficients are given in Table~\ref{tab:v20_ebop}, 
{and $(O-C)$ residuals are shown in Figures~\ref{fig:v20_oc_v} and 
\ref{fig:v20_oc_i},
which clearly reveal the higher quality of the $V$ light curve.
Except for a larger scatter, partly due to small nightly differences, there
are no serious systematic trends in the $I$ residuals.}
As seen, the orbital inclination is close to 90{\degr}, and 
the elements of the individual solutions agree well.
The two quite different components are small compared to the radius of the
orbit; for both components the individual radii obtained agree within
about $\pm1\%$. Additional tests reveal that the ratio of relative radii
$k$ is well constrained by the light curves.

\input 0749tab3.tex

\input 0749tab4.tex

\input 0749tab5.tex

Including the linear limb darkening coefficients for both components
as adjustable parameters was only partly successful, since those for the
secondary components converged towards unrealistically low and also
very uncertain values. For the primary component, the $V$ value is
close to that by Van Hamme (\cite{vh93}), whereas the $I$ value is
just between the two theoretical coefficients.
They are almost independent of whether $u_s$ is fixed at the theoretical
values or adjusted.  

Adopting a non-linear limb darkening law (square root) was also attempted
but has no significant influence on the solutions, as also found by
Lacy, Torres \& Claret (\cite{lacy08}).

As already mentioned the amount of third light is not well constrained
by the light curves themselves. If included as free parameters, much
too low but also very uncertain values are obtained
($V: 0.04 \pm 0.08; I: 0.13 \pm 0.10$). 
Upper third light limits, leading to an orbital
inclination of 90{\degr}, are about 0.16 ($V$) and 0.18 ($I$).
The effect on the photometric elements of changing the adopted $l_3$
values by 15\%, i.e. by about their uncertainties given above, 
is shown in Table~\ref{tab:v20_ebop_l3}.

The final, adopted photometric elements for V20 are listed in 
Table~\ref{tab:v20_phel}. Higher weight has been given
to the results based on Van Hamme (\cite{vh93}) linear
limb darkening coefficients.
Uncertainties are based on interagreement between the $V$ 
and $I$ solutions, the uncertainties of the amount of third 
light, and Monte Carlo simulations.

\section{Spectroscopy for V20}
\label{sec:spec}

 The spectroscopic observations for V20 were carried out in service mode
 with UVES at the ESO VLT during allocation periods 75 and 77. Since V20
 is at declination $+37\degr$, it can on Paranal only be observed at an 
airmass larger than 2.1 -- therefore all observations were carried out 
near meridian passage.
 Due to the faintness of V20, and in order to minimize slit losses, a slit
 of 1\farcs20 width, corresponding to a resolution of approximately
 37\,000, was used. The slit was aligned along the parallactic angle
 (ELEV mode), and the ADC was not inserted in the beam, since it causes a
 slight loss of flux.  
 The standard 580nm setup, and on-chip binning of 2$\times$2 pixels,
was used for all observations. 
 The corresponding wavelength ranges covered at the two CCD detectors 
employed in UVES are approximately 4775-5750{\AA} and 5875-6830{\AA},
respectively.
A total of 17 epochs were obtained, see Table~\ref{tab:v20_UVES_log}, each
with a ThAr exposure attached.
The FWHM from the image headers for the start and end of each exposure 
are included.
The typical S/N per pixel for the red chip was between 15 and 25; 
two of the spectra had to be omitted in the data analysis due to
a very low signal. After careful check of the wavelength calibrations, we 
decided to apply the pipeline reduced spectra for the analyses described below.

\subsection{Radial velocities and spectroscopic elements}
\label{sec:rv}

In order to reduce/eliminate possible systematic velocity errors, we
have decided to apply the broadening function (BF) formalism 
(Rucinski \cite{r99,r02,r04}; with useful IDL routines and a 
cookbook from his 
homepage\footnote{{\scriptsize\tt http://www.astro.utoronto.ca/$\sim$rucinski/}}) 
for radial velocity measurements of V20.
Based on results reported in the literature, we expected this
method to be well suited for the analysis of the sharp lined but
otherwise non-trivial spectra of V20 with three set of lines. 
As seen below, our expectations have been fully met.

\input 0749tab6.tex

\begin{figure}
\epsfxsize=80mm
\epsfbox{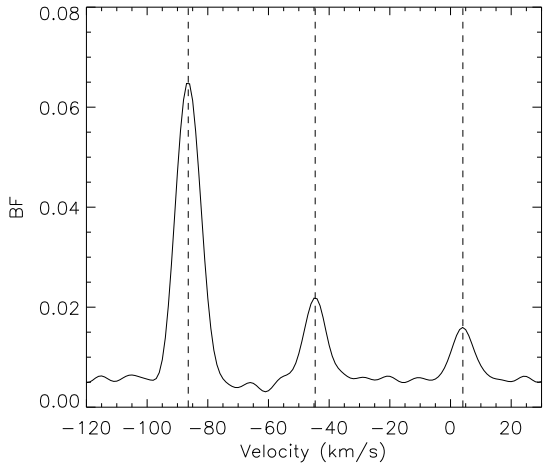}
\caption[]
{
\label{fig:v20_bf}
Broadening function obtained for the 6100-6700\,\AA\ region
using the primary (P) template.
The UVES spectrum was taken at phase 0.342, at HJD$=$2453590.65270.
From left to right the components are: primary, third star, secondary.
}
\end{figure}

\input 0749tab7.tex

\begin{figure}
\epsfxsize=90mm
\epsfbox{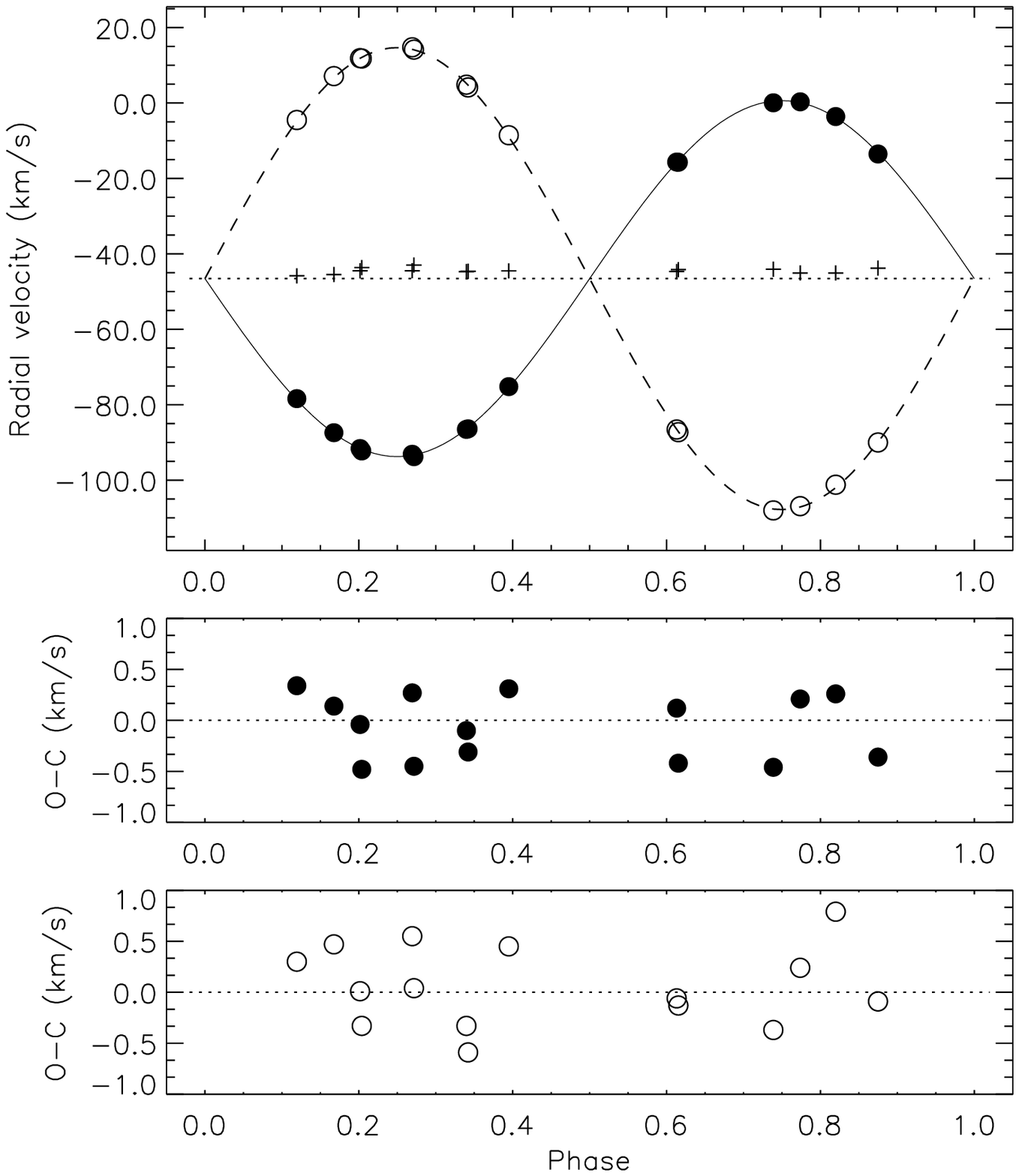}
\caption[]
{
\label{fig:v20_orbit}
Spectroscopic double-lined orbital solution for V20 obtained from
the 6100-6700 {\AA} region using the primary (P) template;
see Table~\ref{tab:v20_spel}.
Filled circle: primary; open circle: secondary.
The horizontal dotted line (upper panel) represents the center--of--mass
velocity of V20. The $+$ signs are the measured velocities for the
third light component.
Phase 0.0 corresponds to central primary eclipse.
}
\end{figure}

\input 0749tab8.tex   

The BF method applies only one (unbroadened) template at a time, 
so we have decided to perform several analyses adopting three different
synthetic templates, corresponding to each of the three set of lines
present in the V20 spectra.
For calculation of templates we have used the $bssynth$ tool
(Bruntt, private communication), which applies the SYNTH software
(Valenti \& Piskunov \cite{vp96}) and modified ATLAS9 models
(Heiter \cite{heiter02}).
The UVES spectra were first carefully cleaned and normalized, and
then, like the template spectra, logarithmically rebinned to a 
constant velocity step of 1.0 \kms.  

The two spectral regions were analysed separately, and
the position of the BF's, defining the radial velocities of the stars, 
were calculated by fitting Gaussians to the smoothed functions. 
A sample BF is shown in Fig.~\ref{fig:v20_bf}.

Spectroscopic elements were derived from the measured radial velocities 
using the method of Lehman-Filh\'es implemented in the SBOP program 
(Etzel \cite{e04}). The orbital period given in Eq.~\ref{eq:v20_eph} was 
adopted, and the orbit was defined to be circular. Both double-lined
and single lined solutions were done. Assuming that the radial
velocity of the third component is constant, small shifts corresponding
to the difference between its mean velocity and the individual
values were added to measured velocities of the primary and
secondary components. The shifts do not alter the derived 
system velocities and semi-amplitudes significantly but improve 
their uncertainties.

The orbital elements determined from the
different spectral ranges using different templates are given
in Table~\ref{tab:v20_spel} together with the mean radial velocities
for the third star. As seen, very accurate velocity semi-amplitudes
are obtained; in general, the highest accuracy is reached for the
6100-6700 {\AA} region; Fig.~\ref{fig:v20_orbit} illustrates one
of the solutions.
The double-lined solutions and the single-lined solutions from the 
6100-6700 {\AA} region agree extremely well, whereas the semi-amplitudes 
of the single-lined solutions from the 5100-5600 {\AA} region are 
systematically lower by about 0.10 \kms.
The primary and secondary single-lined solutions, except the
secondary for the 5100-5600 {\AA} region, give slightly
different values for the system velocity, although the results
formally agree within the errors. 

In order to check for possible systematic velocity errors, we
have analysed synthetic spectra calculated for the observed orbital
phases. The three templates, broadened to $v$sin$i$ values 
matching the observations, were shifted to their velocities at a
given phase and combined according to the relative luminosities of the 
V20 components and the third star.
BF analyses of the synthetic spectra show that the measured radial
velocities deviate only slightly from the input values, typically
below $\pm 0.15$ \kms\ for the primary and $\pm 0.30$ \kms\ for the 
secondary. Furthermore, adding the small shifts to the radial velocities 
measured from the observed spectra does not lead to significant changes of
the orbital elements and their uncertainties.
Two examples (marked by $*$) are included in Table~\ref{tab:v20_spel}.

We adopt the weighted mean of the uncorrected single-lined solutions as the 
final spectroscopic elements for V20 listed in Table~\ref{tab:v20_orbit}.
Errors typical for the individual solutions have conservatively been
assigned to the semi-amplitudes and the system velocity.
As seen, minimum masses accurate to 0.7\% and 0.5\% have been obtained
for the primary and secondary components, respectively.
For comparison, masses derived from the corrected single-lined solutions 
are only 0.2\% higher for the primary and 0.2\% lower for the secondary 
component.

We have determined a systemic velocity for V20 of 
$-46.63 \pm 0.13$ \kms (Table \ref{tab:v20_orbit}). 
This is in excellent agreement
with the value of $-47.1 \pm 0.8$ \kms as determined from 
15 cluster members by Carraro et al.  (\cite{carraro06}) 
who also found the dispersion in the radial velocities of the 15 stars
to be $2.2 \pm 0.4$ \kms.
This leaves little doubt that the system is indeed a 
member of NGC\,6791. 

\section{Absolute dimensions and distance for V20}
\label{sec:absdim}

\input 0749tab9.tex

Absolute dimensions for the components of V20 are calculated
from the elements given in Tables~\ref{tab:v20_phel} and
\ref{tab:v20_orbit}.
As seen in Table~\ref{tab:v20_absdim}, masses and radii have been
obtained to an accuracy of about 0.6\% and 1.0\%, respectively.
Individual $V,I,V-I$ magnitudes and indices are included, 
as calculated from the combined
photometry (Table~\ref{tab:v20_VI}) and the luminosity ratios
between the components (Table~\ref{tab:v20_phel}). An uncertainty of
15\%, correlated in $V$ and $I$, has been assumed for $l_3$.

For the determination of effective temperatures from the $(V-I)$
indices, we adopt an interstellar reddening of
$E(B-V) = 0.15 \pm 0.02$, see Sect.~\ref{sec:reddening}, and assume
[Fe/H] = $+0.40 \pm 0.10$, see Sect.~\ref{sec:metallicity}.
The calibration by Ram\'\i rez \& Mel\'endez (\cite{rm05})
then gives $5715 \pm 125$ K and $4750 \pm 150$ K for the primary
and secondary components, respectively, 
whereas $5665 \pm 100$ K and $4900 \pm 100$ K
are obtained from the recent calibration by Casagrande et al. (\cite{cas06}).
From the empirical flux scale by Popper (\cite{dmp80}), the
bolometric correction (BC) scale
by Flower (\cite{flower96}), and the $V$ flux ratio between the components, 
which is obtained to high precision from the light curve analyses 
(Table~\ref{tab:v20_phel}), we get a well constrained temperature 
difference of $765 \pm 15$ K, supporting the results from the
Casagrande et al. calibration.
We consequently adopt 5665 and 4900 K as our final $T_{\rm{eff}}$ results.

As seen in Table~\ref{tab:v20_absdim} the synchronous rotational velocities
of the components of V20 are small. Synthetic binary spectra based on these
velocities and a resolution of 37\,000 qualitatively agree well with the
observed spectra, but we have not attempted to determine the actual
rotational velocities.

Eclipsing binaries are important primary distance indicators
(Clausen \cite{clausen04}), and V20 provides a more direct determination
for NGC\,6791 than traditional main-sequence fitting. 
The two components yield nearly identical distances 
(4000 pc or $(V_0 - M_V) = 13\fm00$), 
which have been established to a robust accuracy of 5\%, 
taking into account all error sources.
For comparison, recently published results from main-sequence fitting,
which show a spread significantly higher than the quoted errors, 
are: 
$12\fm79$ (SBG03; $E(B-V)$ = 0.09 and [Fe/H] = 0.3), 
$13\fm07 \pm 0\fm04$ (Carney et al.  \cite{carney05}; $E(B-V)$ = 0.14 and [Fe/H] = 0.4), 
and $13\fm1 \pm 0\fm1$ (ATTM07; $E(B-V)$ = 0.155 and [Fe/H] = 0.45).

\section{NGC\,6791}
\label{sec:ngc6791}
\subsection{The reddening of NGC\,6791}
\label{sec:reddening}

 Anthony--Twarog, Twarog \& Mayer (\cite{attm07}) gave a thorough
 discussion of the reddening towards NGC\,6791 based on $vbyCaH\beta$ photometry
 and a recalibration suitable for metal-rich stars, such as those found in 
 this cluster. Over the years, since the first study of Kinman 
(\cite{kinman65}), there has been significant disagreement over the reddening 
value. Here we adopt a value of $E(B-V)=0.15\pm0.02$ for the cluster, 
giving most weight to the ATTM07 value which also agrees well with the 
value derived from the maps of Schlegel et al. (\cite{sch98}).  

\subsection{Cluster metallicity}
\label{sec:metallicity}

 As is the case for the reddening towards NGC\,6791, the determination
 of its metallicity also has a long history. We shall not repeat it 
 here but refer the reader to ATTM07 and Carretta et al. (\cite{carretta07})
for a more comprehensive discussion. 
 Based on the intense interest over the past few years, we shall 
 adopt the values determined spectroscopically. While a full 
 consensus has not been reached on this subject, most recent 
 determinations seem to agree on a value close to \feh\,$=+0.40$,
which we adopt together with an uncertainty of $\pm0.10$ dex. 

We note that e.g. Carretta et al. report significant underabundances 
for C,N, and O of about -0.3 dex. Like Carraro et al. (\cite{carraro06})
they find scaled-solar $\alpha$-element abundances.

\section{Comparison with theoretical models}
\label{sec:models}

In the following, we compare the accurate dimensions obtained for
V20, and the NGC\,6791 CMDs, with properties of the VRSS 
(scaled-solar mixture) Victoria-Regina evolutionary tracks and 
isochrones (VandenBerg et al., 
\cite{vr06})\footnote{{\scriptsize\tt http://www1.cadc-ccda.hia-iha.nrc-cnrc.gc.ca/
cvo/community/VictoriaReginaModels/}}, 
the $Y^2$ (Yonsei-Yale) grids
(Demarque et al. \cite{yale04})\footnote{{\scriptsize\tt http://www.astro.yale.edu/demarque/yystar.html}}, 
and the extensive material available from the BaSTI database
(Pietrinferni et al., \cite{basti04})\footnote{{\scriptsize\tt http://www.te.astro.it/BASTI/index.php}}.  
These recent models are all based on up-to-date input physics but differ 
somewhat with respect to e.g. core overshoot and diffusion (if included)
treatment, He enrichment law, adopted solar mixture, and envelope 
convection calibration. Furthermore their associated colour transformations 
are different.
We refer to Clausen et al. (\cite{avw08}) for further details, as well as
to the original papers. 

Throughout we have adopted scaled-solar mixture for the heavy elements, and
we have selected the available models and isochrones which are closest to the 
adopted \feh = $+0.40 \pm 0.10$ for the comparisons.

\subsection{V20}
\label{sec:models_v20}

For a given mass and \feh, the observable properties
of models at a given age like radius, effective temperature, and
luminosity depend on the adopted input physics, including treatment
of core and envelope convection, diffusion etc., and the assumed 
Y,Z relation. 
In order to constrain such free model parameters,
accurate results from many binaries are needed; cluster members are
particularly valuable.
From the binary perspective, masses and radii are the most direct
parameters available, free of any scale dependent calibrations.
So, the $M-R$ diagram is well suited for isochrone tests,
especially when the binary components are significantly different,
as is the case for V20.
In addition, the $T_{\rm{eff}}-R$ plane allows tests of model 
temperatures, which for a given mass and \feh\ depend on e.g. 
abundance mixture leading to Z, on Y, and on surface 
convection efficiency.
Adopting the usual assumption of coeval formation of the components 
from the same raw material, identical ages must be reached in 
these two planes, as well as in the $M-\mathrm{log}(L)$ plane.

\input 0749ta10.tex

\begin{figure}
\epsfxsize=90mm
\epsfbox{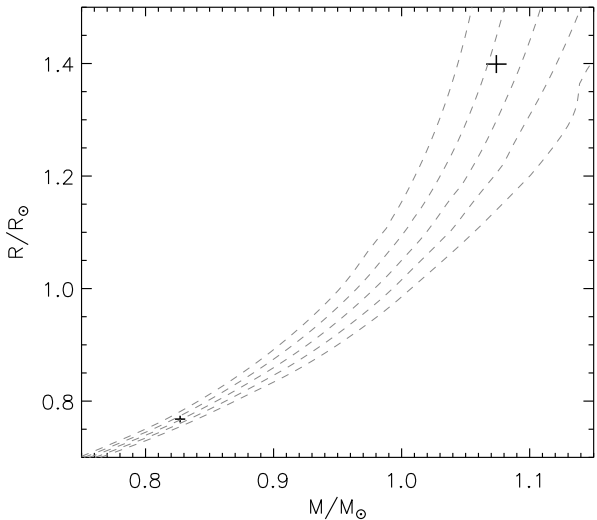}
\caption[]{\label{fig:v20_mr_vic}
V20 compared to Victoria-Regina VRSS models for
($X$,$Y$,$Z$) = (0.63656,0.32344,0.04000), equivalent to
\feh\,$=+0.37$ for \afe\,$=0.00$.
Isochrones from 5.0 to 9.0 Gyr (step 1.0 Gyr) are shown.
}
\end{figure}

\begin{figure}
\epsfxsize=90mm
\epsfbox{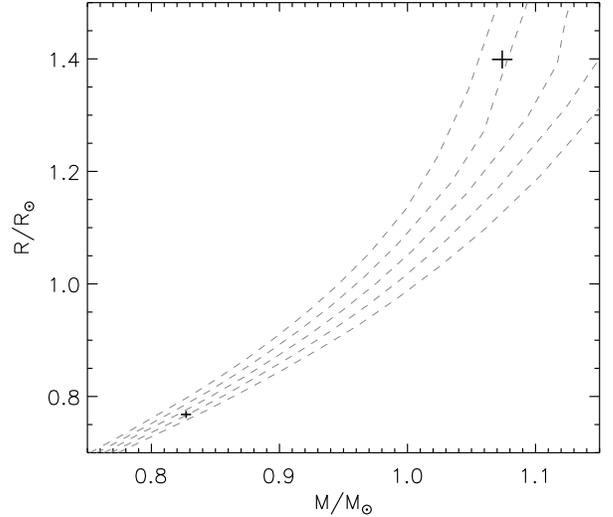}
\caption[]{\label{fig:v20_mr_yal040}
V20 compared to $Y^2$ models for
($X$,$Y$,$Z$) = (0.6467,0.3122,0.0411), equivalent to
\feh\,$=+0.40$ for \afe\,$=0.00$.
Isochrones from 5.0 to 9.0 Gyr (step 1.0 Gyr) are shown.
}
\end{figure}

\begin{figure}
\epsfxsize=90mm
\epsfbox{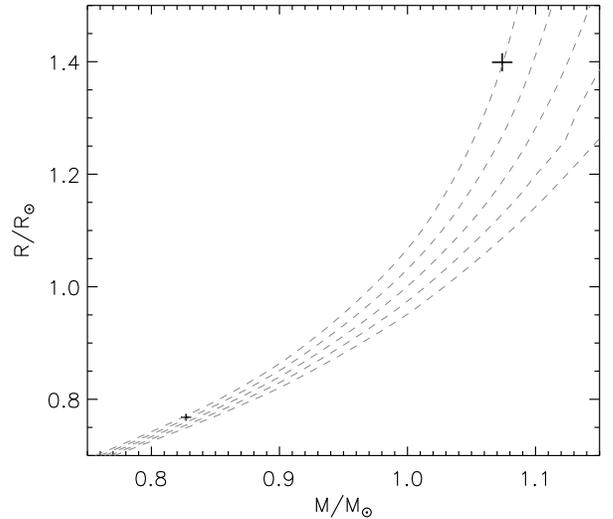}
\caption[]{\label{fig:v20_mr_bas_ov}
V20 compared to BaSTI models with overshooting for
($X$,$Y$,$Z$) = (0.657,0.303,0.0400), equivalent to
\feh\,$=+0.395$ for \afe\,$=0.00$.
Isochrones from 5.0 to 9.0 Gyr (step 1.0 Gyr) are shown.
}
\end{figure}

From the $M-R$ comparisons shown in Figs.~\ref{fig:v20_mr_vic},
\ref{fig:v20_mr_yal040}, and \ref{fig:v20_mr_bas_ov} we derive
the component ages listed in Table~\ref{tab:v20_age}. 
As seen, mass and radius uncertainties translate to age uncertainties 
of about 0.3 Gyr (primary) and 0.9 Gyr (secondary), respectively, for a 
given adopted composition. However, at \feh = 0.40, the ages
resulting from the three grids range between 7.6 and 9.0 Gyr for
the primary and between 6.2 and 8.6 for the secondary.
Within uncertainties, the Victoria-Regina and BaSTI isochrones predict
identical ages for the two components, although the age for the
secondary is systematically lower than that of the primary.
The $Y^2$ grids clearly predict different ages, again with lower
values for the secondary. 
Age uncertainties due to an \feh\, uncertainty of 0.1 dex
are comparable to those coming from masses and radii.

We note that this trend of lower
ages for the less massive star is opposite to what has been
found for several field G-type eclipsing binaries; see e.g.
Popper (\cite{dmp97}), Clausen et al. (\cite{granada99}), 
Torres et al. (\cite{wt06}), and Clausen et al. (in prep.). 
In these cases, the seemingly higher ages may, 
although not yet firmly proved, be due to surface activity 
and correlated less efficient convection at the less massive 
components, which result in larger radius and lower effective 
temperature (but nearly unchanged luminosity).

\begin{figure}
\epsfxsize=90mm
\epsfbox{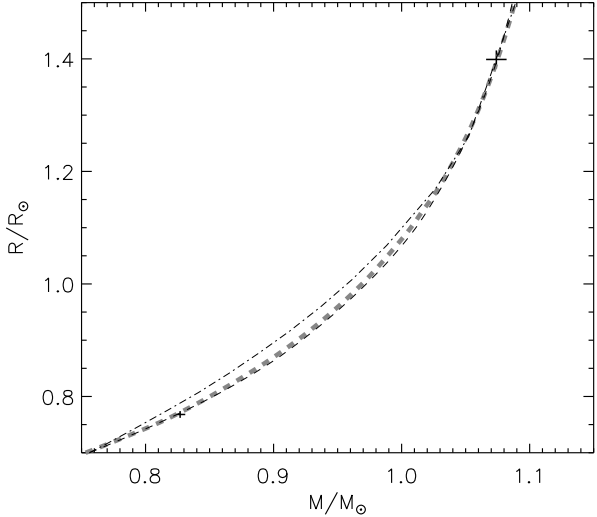}
\caption[]{\label{fig:v20_mr_comp}

V20 compared to the VRSS (7.7 Gyr, dashed thick gray), 
$Y^2$ (8.2 Gyr, dashed-dot black), 
and BaSTI (9.0 Gyr, dashed black) isochrones which fit the primary 
component at the observed [Fe/H]; 
see Table~\ref{tab:v20_age}.
}
\end{figure}

In Fig.~\ref{fig:v20_mr_comp}, we compare the Victoria-Regina,
BaSTI, and $Y^2$ isochrones fitting the primary component.
The first two are nearly identical, but for an age difference of
as much as 1.3 Gyr, whereas the shape of the $Y^2$ isochrone,
at an age approximately between the two, is significantly different.

\begin{figure}
\epsfxsize=90mm
\epsfbox{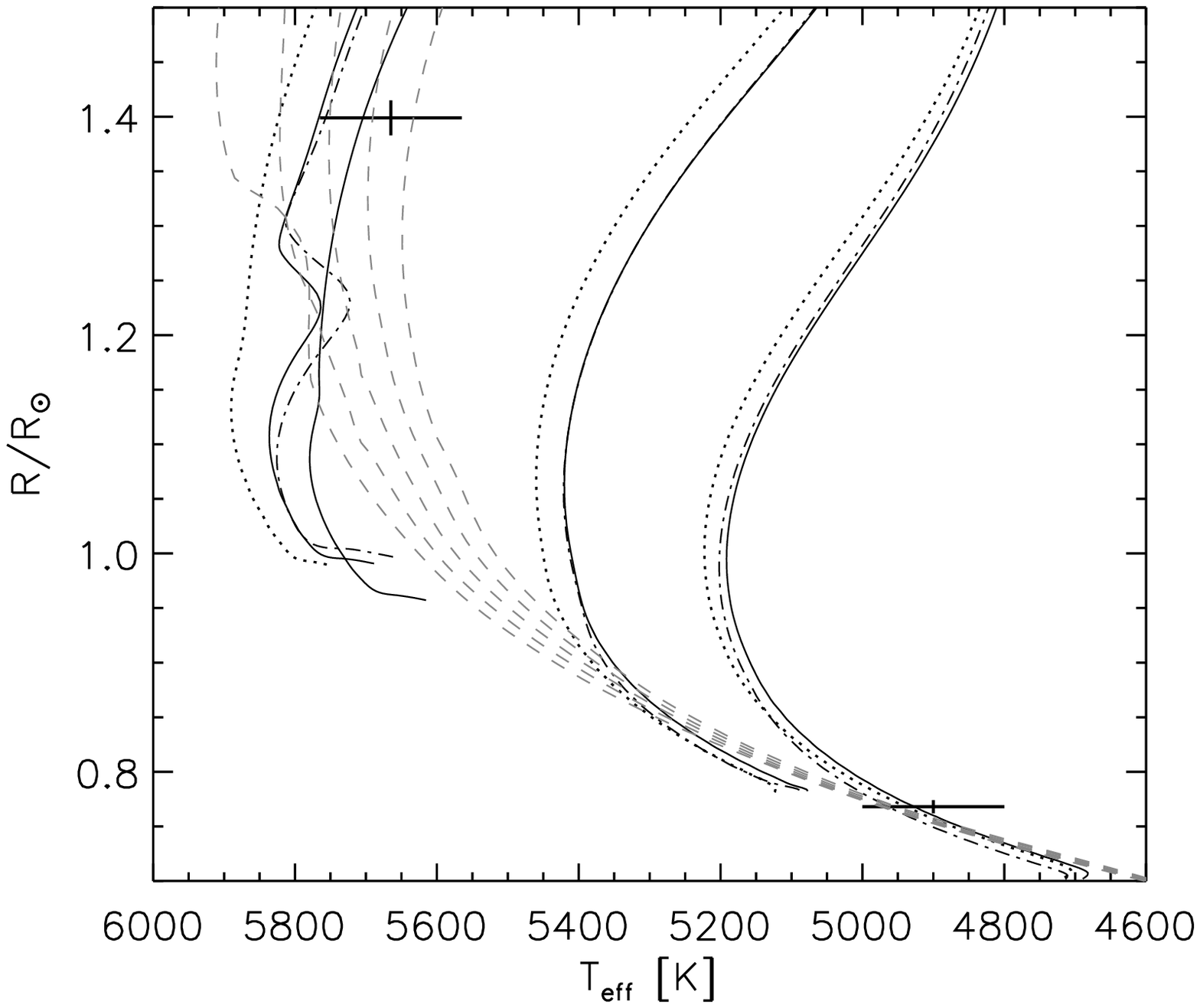}
\caption[]{\label{fig:v20_tr_vic}
V20 compared to Victoria-Regina VRSS models for
($X$,$Y$,$Z$) = (0.63656,0.32344,0.04000), equivalent to
\feh\,$=+0.37$ for \afe\,$=0.00$.
Tracks for 0.8, 0.9, 1.073, and 1.1 $M_{\sun}$ (full drawn) and
isochrones from 5.0 to 9.0 Gyr (dashed, step 1.0 Gyr) are shown.
For comparison, tracks for 0.8, 0.9, and 1.1 $M_{\sun}$ for 
($X$,$Y$,$Z$) = (0.66856,0.30144,0.03000), 
equivalent to \feh\,$=+0.23$ for \afe\,$=0.00$ (dotted), and
for ($X$,$Y$,$Z$) = (0.60456,0.34544,0.05000), 
equivalent to \feh\,$=+0.49$ for \afe\,$=0.00$ (dot-dash), 
are included.
}
\end{figure}

\begin{figure}
\epsfxsize=90mm
\epsfbox{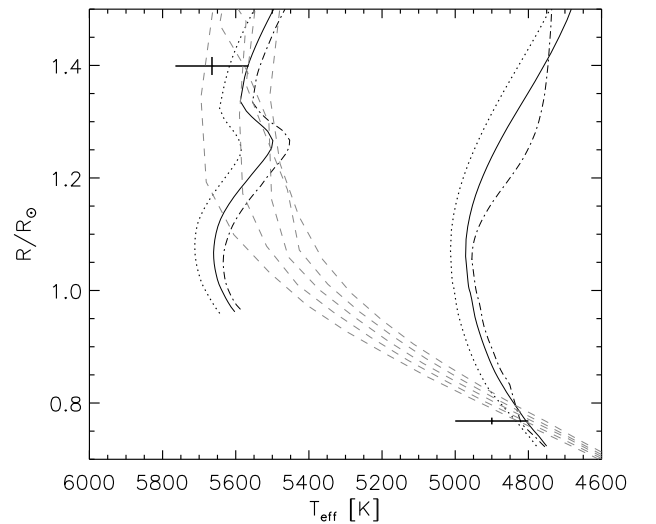}
\caption[]{\label{fig:v20_tr_yal040}
V20 compared to $Y^2$ models for
($X$,$Y$,$Z$) = (0.6467,0.3122,0.0411), equivalent to
\feh\,$=+0.40$ for \afe\,$=0.0$.
Tracks for the component masses (1.074 and 0.827 $M_{\sun}$, full drawn) and
isochrones from 5.0 to 9.0 Gyr (dashed, step 1.0 Gyr) are shown.
To illustrate the effect of the abundance uncertainty, tracks for 
\feh\,$=+0.30$ and \afe\,$=0.00$ (dotted), corresponding to 
($X$,$Y$,$Z$) = (0.6686,0.2976,0.0338), 
and \feh\,$=+0.50$ and \afe\,$=0.00$ (dash-dot), corresponding to
($X$,$Y$,$Z$) = (0.6200,0.3300,0.0500),
are included.
}
\end{figure}

\begin{figure}
\epsfxsize=90mm
\epsfbox{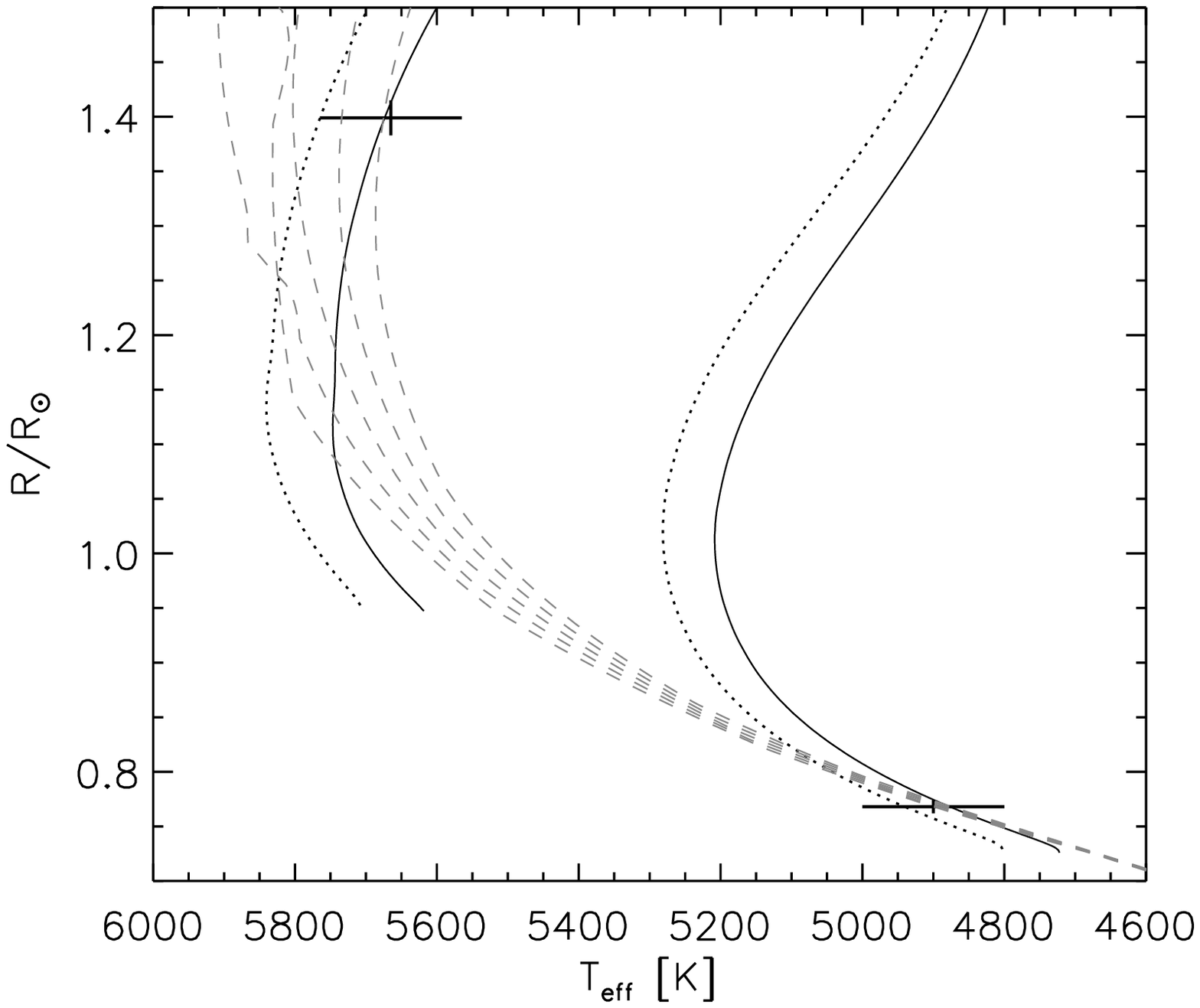}
\caption[]{\label{fig:v20_tr_bas_ov}
V20 compared to BaSTI models with overshooting for
($X$,$Y$,$Z$) = (0.657,0.303,0.0400), equivalent to
\feh\,$=+0.395$ for \afe\,$=0.00$.
Tracks for the component masses (1.074 and 0.827 $M_{\sun}$, full drawn) and
isochrones from 5.0 to 9.0 Gyr (dashed, step 1.0 Gyr) are shown.
For comparison, tracks (dotted) for ($X$,$Y$,$Z$) = (0.682,0.288,0.0300), 
equivalent to \feh\,$=+0.254$ for \afe\,$=0.00$, are included.
}
\end{figure}

Turning to the $T_{\rm{eff}}-R$ plane shown in Figs.~\ref{fig:v20_tr_vic},
\ref{fig:v20_tr_yal040}, and \ref{fig:v20_tr_bas_ov}, it is seen
that the Victoria-Regina tracks fit both components within errors.
They are very slightly hotter than observed, and as seen 
in Fig.~\ref{fig:v20_tr_vic} this is also the case for tracks
calculated for a higher \feh = 0.49, which actually fall close to the
\feh = 0.37 tracks. 
The $Y^2$ tracks are cooler than observed, and here tracks for a 0.1 dex
lower \feh\ agree better.
The BaSTI tracks, both standard and with core overshoot, fit the components 
of V20 perfectly well.

\begin{figure}
\epsfxsize=90mm
\epsfbox{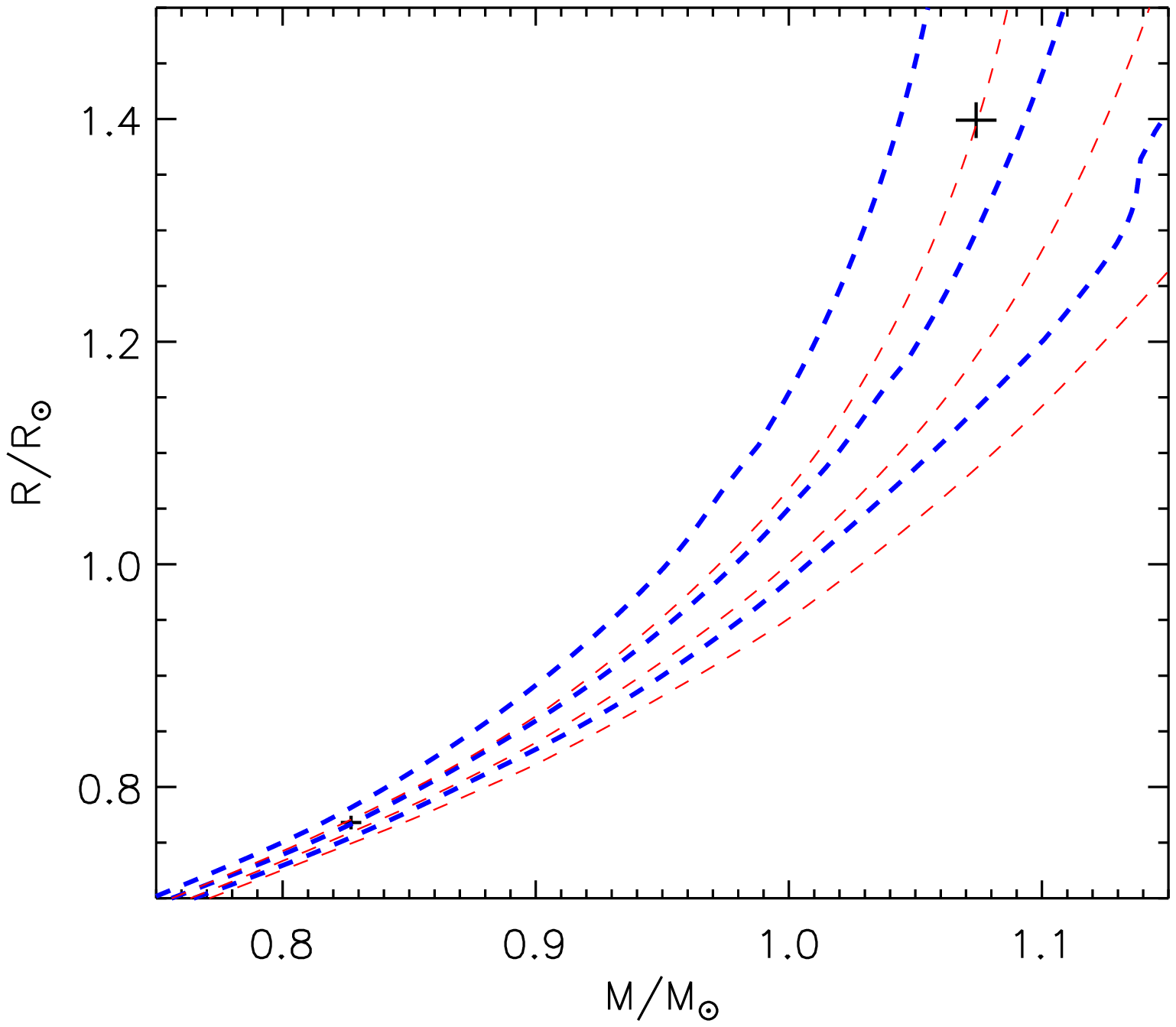}
\caption[]{\label{fig:v20_mr_isocomp}
Comparison between VRSS (thick blue) and BaSTI (thin red) 
isochrones at 5,7, and 9 Gyr for compositions close to the observed [Fe/H]; 
see Table~\ref{tab:v20_age}. The components of V20 are included
for comparison
}
\end{figure}

\begin{figure}
\epsfxsize=90mm
\epsfbox{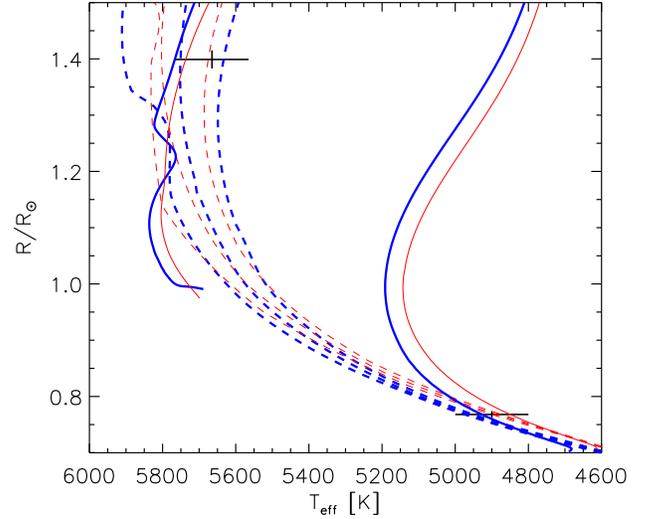} 
\caption[]{\label{fig:v20_tr_lin_comp}
Comparison between VRSS (thick blue) and 
BaSTI (thin red) 1.1 and 0.8 $M_{\sun}$ 
tracks (full drawn) and 5,7, and 9 Gyr isochrones (dashed).
for compositions close to the observed [Fe/H]; see Table~\ref{tab:v20_age}. 
The components of V20 are included for comparison
}
\end{figure}

In conclusion, both the the Victoria-Regina and the BaSTI models
represent V20 well, but at ages which differ by about 1.3 Gyr.
This is significantly higher than the precision of about 0.5 Gyr,
which can be reached from the available mass, radius, and
abundance information.
The cause(s) for this difference is not clear but could perhaps be 
related to the fact, that the Victoria-Regina models apply different
$Y,Z$ relations and core overshoot treatment. 
Direct comparisons between Victoria-Regina and BaSTI tracks and
isochrones are shown in Figs.~\ref{fig:v20_mr_isocomp} and 
\ref{fig:v20_tr_lin_comp}.

Additional NGC\,6791 binaries with component(s) between 0.85 and 
1.0 $M_{\sun}$ would add further constraints on the models, and
would e.g. clearly reveal if the isochrone shape is correct.
Also, similar comparisons for younger clusters would be
valuable. We note that for 1-2 Gyr field F-type binaries with
component masses in the 1.1-1.4 $M_{\sun}$ range, Clausen et al.
(\cite{avw08}) recently found that the Victoria-Regina models
are superior to the BaSTI models.

\subsection{NGC\,6791 colour-magnitude diagrams}\label{sec:cmds_6791}

\begin{figure*}
\centering
\epsfxsize=170mm
\epsfbox{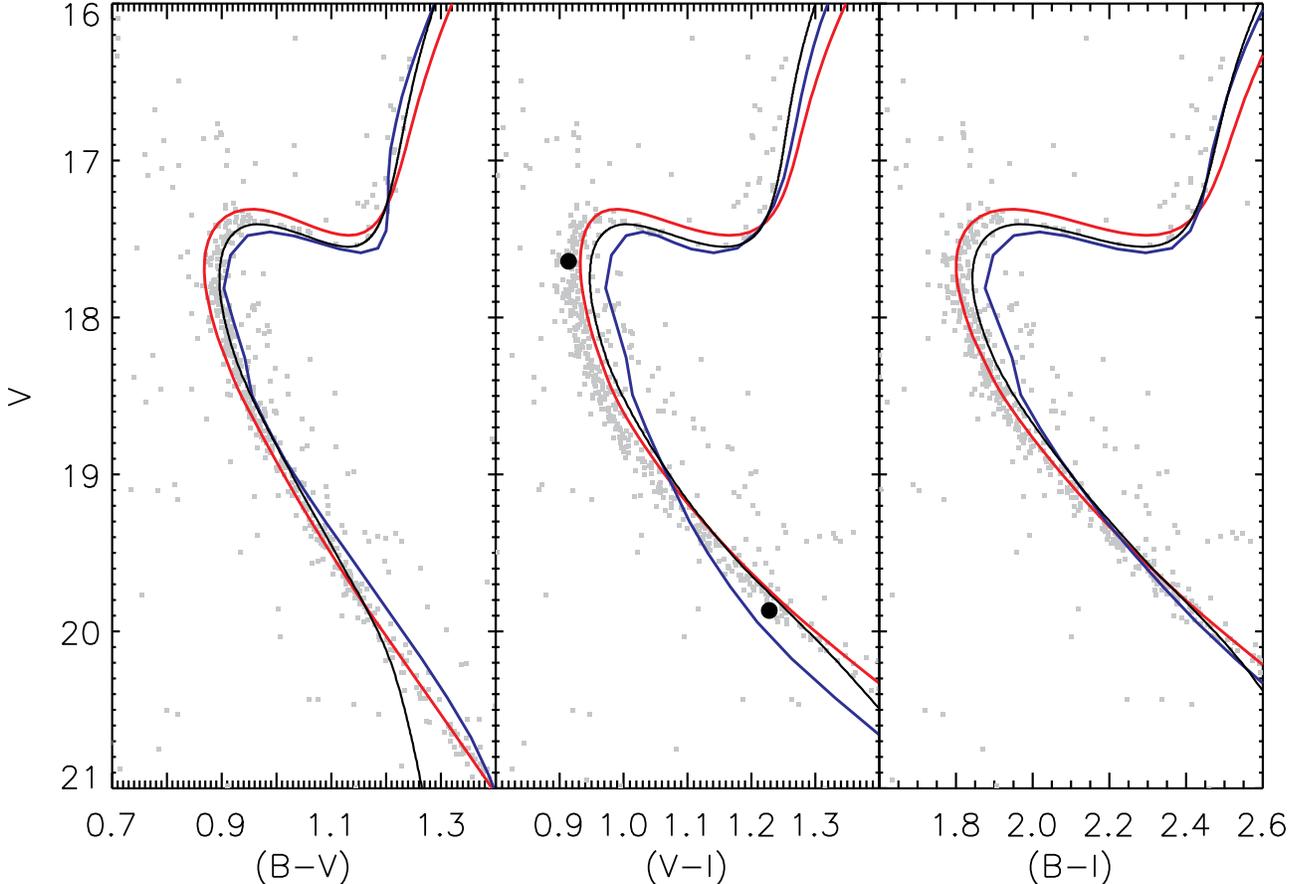}
\caption[]{\label{fig:cmd_compare}
Isochrones for ages 7.7 (VRSS, red), 8.2 ($Y^2$, blue), and 9.0Gyr
(BaSTI, black) overplotted on the various CMDs of NGC\,6791 (SBG03
photometry, plotted at light grey squares). 
 In all plots we have adopted E$(B-V)=0.15$ and $(m-M)_V=13.46$ to
 transpose the isochrones to the observational plane (see
 Table~\ref{tab:v20_absdim}). In the middle panel,
 the location of the primary and secondary component of V20 are shown as
 filled black circles. The models are the same as those used in
 Fig.~\ref{fig:v20_mr_comp}. On black and white printers the models
 are hard to distinguish, but at the cluster turnoff the VRSS models are
 always the bluest and the $Y^2$ models always the reddest. }
\end{figure*}

 In light of the results obtained from V20 in the previous 
 section, it is interesting to investigate what extra information 
 can be gained from a comparison of the isochrones to the CMD 
 of the cluster. There is an abundant literature on age 
 determination of old open-- and globular clusters using 
 ``isochrone-fitting''. The method is well described and so 
 are the problems involved in the process. 
 Observationally it is necessary to obtain accurate determinations 
 of reddening, distance and cluster abundance, and subsequently 
 match theoretical models shifted in colour and luminosity to the 
 observed CMD. The theoretical models must be transformed to the 
 observational plane (colour and magnitude) in order to allow the 
 comparison. It is well known that for old clusters, the isochrone 
 shapes only change slowly with time, and thus age-determinations 
 based on isochrone shapes are not very sensitive. Furthermore, the 
 turnoff luminosity and colour also changes slowly with time. In the 
 best cases, the accuracies obtained for reddening and distance 
 modulus of clusters is of the order $0\fm01$ and $0\fm1$, 
 respectively (Pasquini et al. \cite{pasquini08}). 
 For an 8\,Gyr old cluster such errors 
 translate into an uncertainty in age (from the turnoff position) 
 of $\sim0.5\,$Gyr and $\sim1.0\,$Gyr, respectively (estimated from VRSS 
 isochrones and for \feh = 0.4). Thus, to 
 obtain age estimates better than 1\,Gyr, very precise values for 
 reddening and distance must be obtained -- in addition to this 
 the models also show a non-negligible dependence on the adopted 
 metallicity. 

 VandenBerg \& Stetson (\cite{vdb04}), Fig. 7, present a systematic 
 investigation of what effect plausible changes in metallicity and 
 distance modulus (at fixed reddening) has on the derived age for 
 \object{NGC~188}, which is $\sim$2Gyr younger than NGC\,6791.  Their exercise 
 shows that ages between 5.9 and 8.1 Gyr are compatible with the 
 cluster CMD corresponding to $\sim30\%$ of the cluster age. 

 To confront the theoretical models with the observations we show, in 
 Fig.~\ref{fig:cmd_compare}, the CMD for NGC\,6791 in the $(B-V,V), 
 (V-I,V)$ and $(B-I,V)$ planes with the preferred age for each of the
 three isochrone sets (using the reddening and distance from 
 Table~\ref{tab:v20_absdim}). 
 We employ the models closest to \feh\,$=+0.40$ throughout. 
 It is evident that the isochrones are not far from matching the CMD in 
 all colour planes, although the $Y^2$ models are somewhat too red compared 
 to the VRSS and BaSTI models. From inspection of the figure, it is clear that
 offsets within $0\fm02$ in reddening and $0\fm10$ in distance-modulus 
 would bring the models into agreement with the cluster turnoff region. 
 We thus conclude that for the three age values, the isochrones in the 
 CMD are not in direct conflict with the observations, except perhaps for
 the $(V-I)$ and $(B-I)$ colours of the $Y^2$ models. 
 By simple experimenting with VRSS models it is also clear that with 
 (plausible) modifications to the adopted reddening and distance 
 modulus, it is posible to match the turnoff region for the cluster 
 for ages spanning at least the 6-9 Gyr range; the same conclusion
 would also hold for BaSTI or $Y^2$ models. 

 Given the current uncertainties in the derivation of cluster reddening, 
 distance modulus, and metallicity, it is the opinion of the authors that
 we can only gain limited information on cluster absolute ages from their
 colour-magnitude diagrams. From the results presented here, 
 V20 provide much tighter constraints on the age of NGC\,6791 than 
 isochrone-fitting to the colour-magnitude diagram. 
 In fact, it would be necessary to know the distance-modulus and  
 intrinsic colour of the cluster turnoff to an accuracy better 
 than $0\fm05$ and $0\fm01$, respectively to match the 
 0.5Gyr age precision. 

 Unfortunately, we are faced with the situation that even in the 
 $M-R$ diagram, the models give very discrepant ages with differences
 of 15--20\% (see Fig.~\ref{fig:v20_mr_vic}). Until this situation is 
 clarified, it is the models which limit the precision with which stellar ages 
 can be obtained. We speculate that in the case of NGC\,6791 the discrepancy
 could be due to difficulties in the description of convective energy 
 transport. 

\section{Summary and conclusions}
\label{sec:concl}

 In this paper we have presented extensive photometric and spectroscopic 
 observations of the detached eclipsing binary V20 in the old open 
 cluster NGC\,6791, and determined precise masses and radii for its 
 primary and secondary components. The errors in masses and radii are
 not subject to the usual problems of determining the cluster reddening, 
 metallicity and distance associated with cluster studies.
 This has allowed a determination 
 of the cluster age with an uncertainty of only 0.5Gyr (errors due to 
 mass, radius and metallicity), by comparing the location of the 
 primary component to isochrones in the mass--radius plane. 
 However, the absolute age(s) for the cluster derived from the 
 Victoria-Regina, BaSTI and $Y^2$ isochrones differ by up to 1.3Gyr, 
 and yet the two extreme isochrones overlap almost perfectly in a 
 $M-R$ diagram (Fig.~\ref{fig:v20_mr_comp}). Until these discrepancies
 are resolved, this prevents us from drawing a firm conclusion on the 
 true value for the cluster age. 

 The error in the age due to the uncertainty in the mass and radius 
 is only 0.3Gyr for a single isochrone set. This is $\sim3$ times better 
 than what is typically obtained using isochrone--fitting using CMDs.

 By combining the stellar radii with a temperatures established 
 from the photometry, we determined the cluster distance to be 
 4~kpc, with good 
 agreement between the value from each binary component. This was used
 to compare the cluster CMD to isochrones, for an adopted reddening of
 $E(B-V)\,=\,0.15$. We found that for the derived distance and age, the
 three isochrone sets match the observed CMD reasonably well -- however 
 the obtainable age precision is not nearly as good as when using the 
 binary components in a mass-radius diagram. 
 
 Our main conclusion is that with the needed precision of 
 reddening, metallicity and distance estimates, accurate ages of clusters, 
 based on ``isochrone--fitting'', will be very difficult to obtain until 
 GAIA can provide high accuracy parallaxes. 

 In order to improve on cluster age-determinations, many more eclipsing 
 systems in clusters should be investigated (Kaluzny et al. \cite{kaluzny06}). 
 It would be particularly 
 interesting to study clusters with more than one detached 
 system (see eg. Southworth et al. \cite{southworth04}),
 as this would provide more constraints on the shapes of 
 the isochrones in the $M-R$ diagram at fixed age. A set of well 
 studied clusters with good CMDs and accurate masses and radii for 
 several detached eclipsing systems would provide an excellent 
 calibration set against which stellar models can be tested to high 
 accuracy. By extending such studies to globular clusters detached 
 eclipsing binaries could ultimately provide the strongest constraints
 on their ages. We have shown in this study that using 
 8m--class telescopes it is feasible to work at the required level of
 accuracy at $V\,=\,20$ thus making the turnoff stars in the nearest
 globular clusters, and most open clusters, available.

 Finally, it is worthwhile to note, that with further observations 
 of V20, it will be 
 possible to reduce the size of the errorbars for both the mass and 
 radius determination and thus provide even tighter constraints (for a
 given set of models) on the age of NGC\,6791. With currently existing 
 instrumentation it should be possible to reduce the errors in mass and 
 radius to below 0.5\% for both the primary and secondary components.

\begin{acknowledgements}
We thank H. Bruntt for making his {\it bssynth} software available
and J. Southworth for access to JKTEBOP.
P. Stetson is thanked for sharing his excellent photometry software
with us.
The project "Stars: Central engines of the evolution of the Universe",
carried out at Aarhus University and the University of Copenhagen, is supported
by the Danish National Science Research Council.
FG acknowledges financial support from the Carlsberg Foundation, the
Danish AsteroSeismology Centre at Aarhus University, and 
Instrumentcenter for Dansk Astronomi. 
The following internet-based resources were used in research for
this paper: the NASA Astrophysics Data System; the SIMBAD database
and the ViziR service operated by CDS, Strasbourg, France; the
ar$\chi$iv scientific paper preprint service operated by Cornell University.
\end{acknowledgements}

{}

\listofobjects

\end{document}

%% file: 0749tab1.tex
\begin{table}
\caption[]{\label{tab:v20_VI}
Standard $V,I$ photometry for V20 outside eclipses from
SBG03;  
V20 is star number 8600 in
their list. The mean values, with errors, are based on 925 ($V$)
and 703 ($I$) observations, respectively.
Photometry for the two components and the 3rd star 
was calculated from the depths of the total secondary eclipse, 
assuming that all three stars are located on the main sequence 
of NGC6791; see text for details.

}
\begin{flushleft}
\begin{tabular}{lrrr} \hline
\hline
Object                    & $V$       & $I$ &   $V-I$ \\
\hline
V20 + companion (SBG03) & 17.3390 & 16.3560 & 0.9830\vspace{-0.8mm}\\
                          &$\pm   5$&$\pm   6$&$\pm  8$\\
Primary                   & 17.6400 & 16.7277 & 0.9123 \\
Secondary                 & 19.8686 & 18.6353 & 1.2333 \\
3rd star                  & 19.4322 & 18.2911 & 1.1322 \\
\hline
\end{tabular}
\end{flushleft}
\end{table}

%% file: 0749tab2.tex
\begin{table}
\caption[]{\label{tab:v20_tmin}
Times of minima for V20. 
O-C values 
are calculated for the ephemeris given in Eq.~\ref{eq:v20_eph}
adopting a circular orbit.
References are: 
M2002, Mochejska et al. \cite{m02};
B2003, Bruntt et al. \cite{hb03};
G2008, this paper.
} 
\begin{flushleft}
\begin{tabular}{llcrcl} \hline
\hline
HJD           & rms       & Type &   O-C      &  Band & Ref.  \\    
-2400000      &           &      &            &       &            \\    
\hline
52109.77415   &0.00110    & P    &  0.00215   & $R$   & M2002       \\
53151.60485   &0.00030    & P    &$-0.00161$  & $V$   & G2008       \\ 
53151.60449   &0.00100    & P    &$-0.00125$  & $I$   &  -          \\ 
53180.54732   &0.00060    & P    &  0.00153   & $V$   &  -  \\ 
53180.54747   &0.00070    & P    &  0.00138   & $I$   &  -  \\ 
52102.53467   &0.00100    & S    &$-0.00373$  & $V$   & B2003 \\     
52102.53332   &0.00080    & S    &$-0.00238$  & $V$   &  -    \\
52811.56249   &0.00100    & S    &  0.00101   & $V$   & G2008      \\
52811.56404   &0.00090    & S    &$-0.00054$  & $I$   &  -         \\
52869.44066   &0.00210    & S    &  0.00039   & $V$   &  -  \\
52869.44309   &0.00170    & S    &$-0.00204$  & $I$   &  -  \\
\hline
\end{tabular}
\end{flushleft}
\end{table}

%% file: 0749tab3.tex
\begin{table}
\caption[]{\label{tab:v20_ebop}
Photometric solutions for V20 from the JKTEBOP code.
A photometric scale factor (the magnitude at quadrature) 
and the phase of primary eclipse were included as free parameters.
Linear limb darkening coefficients from Claret (\cite{c00}), 
van Hamme (\cite{vh93}), and free, respectively.
The errors quoted for the adjusted parameters are the $formal$ errors 
determined from the iterative least squares solution procedure.
}
\scriptsize{
\begin{flushleft}
\begin{tabular}{lrrrrrr} \hline
\hline\noalign{\smallskip}
Band                 &  $V$       & $V$      & $V$      & $I$     &  $I$    & $I$\\ 
Limb                 &  C00       & VH93     & Free     & C00     &  VH93   & Free \\
darkening            &            &          &          &         &         &      \\
\hline\noalign{\smallskip}            
$i$ \,(\degr)        &  89.76     &  89.74   &  89.85   & 89.78   & 89.78   & 89.86\vspace{-0.8mm}\\
                     & $\pm11$    & $\pm 9$  & $\pm17$  &$\pm16$  & $\pm15$ & $\pm28$\\

$r_p + r_s$          &  0.0703    & 0.0695   & 0.0696   & 0.0709  & 0.0703  & 0.0705\vspace{-0.8mm}\\
                     &  $\pm 3$   & $\pm 3$  &$\pm  4$  & $\pm 5$ & $\pm 5$ &$\pm  5$ \\

$k=r_s/r_p$          &  0.539     & 0.548    & 0.545    & 0.544   & 0.553   & 0.547\vspace{-0.8mm}\\
                     &  $\pm 1$   & $\pm 1$  &$\pm 3$   & $\pm 2$ & $\pm 2$ &$\pm 4$  \\

$r_p$                &  0.0457    & 0.0449   & 0.0450   & 0.0459  & 0.0454  & 0.0456\\

$r_s$                &  0.0246    & 0.0246   & 0.0245   & 0.0250  & 0.0250  & 0.0249\\

$u_p$                &  0.72      & 0.61     &  0.65    & 0.56    & 0.44    & 0.51\vspace{-0.8mm}\\ 
                     &            &          &$\pm 3$   &         &         &$\pm 4$\\

$u_s$                &  0.79      & 0.74     &  0.34    & 0.62    & 0.54    & 0.43\vspace{-0.8mm}\\
                     &            &          &$\pm25$   &         &         &$\pm26$ \\

$y_p$                &  0.39      & 0.39     &  0.39    & 0.25    & 0.25    & 0.25\\

$y_p$                &  0.46      & 0.46     &  0.46    & 0.29    & 0.29    & 0.29\\

$J_s/J_p$            &  0.448     & 0.447    &  0.381   & 0.599   & 0.594   & 0.562\vspace{-0.8mm}\\
                     &  $\pm 3$   & $\pm 3$  & $\pm35$  & $\pm5  $& $\pm 4$ &$\pm 56$ \\

$L_s/L_p$            &  0.126     & 0.127    &  0.127   &  0.173  & 0.174   & 0.174\\

$l_3$                &  0.146     & 0.146    &  0.146   &  0.168  & 0.168   & 0.168\\

$\sigma$(mag.)       &  0.0044    & 0.0044   &  0.0044  &  0.0067 & 0.0067  & 0.0067\\
\noalign{\smallskip}            
\hline
\end{tabular}            
\end{flushleft}            
}                   
\end{table}

%% file: 0749tab4.tex
\begin{table}
\caption[]{\label{tab:v20_ebop_l3}
The effect of changing the amount of third light $l_3$ by 15\%.
Photometric solutions for V20 from the JKTEBOP code.
A photometric scale factor (the magnitude at quadrature) 
and the phase of primary eclipse were included as free parameters.
Linear limb darkening coefficients from van Hamme (\cite{vh93}) were
adopted, see Table~\ref{tab:v20_ebop}.
The errors quoted for the adjusted parameters are the $formal$ errors determined
from the iterative least squares solution procedure.
}
\begin{center}
\begin{tabular}{lrr} \hline
\hline\noalign{\smallskip}
Band                 &  $V$       &      $I$    \\ 
\hline\noalign{\smallskip}        
$i$ \,(\degr)        &  89.63     & 89.62\vspace{-0.8mm}\\
                     & $\pm 6$    & $\pm 9$ \\

$r_p + r_s$          &  0.0697    & 0.0704\vspace{-0.8mm}\\
                     &  $\pm 3$   & $\pm 5$  \\

$k = r_s/r_p$        &  0.540     & 0.542\vspace{-0.8mm}\\
                     &  $\pm 1$   & $\pm 2$  \\

$r_p$                &  0.0453    & 0.0457 \\

$r_s$                &  0.0244    & 0.0248  \\





$J_s/J_p$            &  0.446     & 0.590\vspace{-0.8mm}\\
                     &  $\pm 3$   & $\pm  4$ \\

$L_s/L_p$            &  0.123     & 0.168    \\

$l_3$                &  0.124     & 0.143    \\

$\sigma$ \, (mag.)   &  0.0044    & 0.0067  \\
\noalign{\smallskip}            
\hline
\end{tabular}            
\end{center}            
\end{table}

%% file: 0749tab5.tex
\begin{table}            
\caption[]{\label{tab:v20_phel}
Adopted photometric elements for V20.
The individual flux and luminosity ratios are based
on the mean stellar and orbital parameters.

}
\begin{center}             
\begin{tabular}{llrr}             
\noalign{\smallskip}             
\hline             
\noalign{\smallskip}             
$i$  \,(\degr)   & $89.76  \pm 0.15$   & &\\          
$r_p$            & $0.0452 \pm 0.0005$ & &\\        
$r_s$            & $0.0248 \pm 0.0002$ & &\\        
\end{tabular}             
\begin{tabular}{lrrrr}             
\noalign{\smallskip}             
                 & $V$      & $I$  & & \\           
\noalign{\smallskip}             
$J_s/J_p$        & 0.451    & 0.590 & &\vspace{-0.8mm}\\  
                 & $\pm 7$  &$\pm 8$& &               \\  
$L_s/L_p$        & 0.129    & 0.172 & &\vspace{-0.8mm}\\  
                 & $\pm 4$  &$\pm 6$& &               \\  
\noalign{\smallskip}             
\hline             
\end{tabular}             
\end{center}            
\end{table}

%% file: 0749tab6.tex
{\small
\begin{table}
\caption[]{\label{tab:v20_UVES_log}
Log of spectroscopic observations for V20 obtained with UVES, 
ordered by orbital phase.  The heliocentric 
Date (HJD) is given at mid-exposure, exposure time 
(T$_{\rm exp}$) in seconds, and seeing (FWHM) in arcseconds.
Observations excluded in the analyses are marked by *.}
\begin{flushleft}
\begin{tabular}{lcccc} 
\hline
\hline
       HJD      & Phase          & T$_{\rm exp}$ & Airmass & FWHM  \\
\hline\noalign{\smallskip}
2453847.88977 &  0.11939 & 4060 & 2.34-2.15 & 0.73-0.66 \\
2453573.65870 &  0.16759 & 2700 & 2.20-2.15 & 0.79-0.69 \\
2453588.62071 &  0.20160 & 2700 & 2.19-2.15 & 0.89-0.68 \\
2453588.65279 &  0.20381 & 2700 & 2.15-2.25 & 0.67-0.60 \\
2453589.60199 &  0.26941 & 2700 & 2.27-2.16 & 0.82-0.83 \\
2453589.63401 &  0.27162 & 2700 & 2.16-2.18 & 0.96-0.75 \\
2453590.62070 &  0.33981 & 2700 & 2.18-2.16 & 0.50-0.73 \\
2453590.65270 &  0.34202 & 2700 & 2.16-2.28 & 0.64-0.63 \\
2453880.81608 &  0.39489 & 4060 & 2.24-2.16 & 0.72-      \\
2453623.51512 &  0.61311 & 2700 & 2.23-2.15 & 0.87-0.87 \\
2453623.54710 &  0.61532 & 2700 & 2.15-2.21 & 0.98-0.81 \\
2453986.50872*&  0.69919 & 4060 & 2.37-2.15 &     -1.11 \\  
2453957.58875*&  0.70057 & 4060 & 2.37-2.15 & 1.39-1.95 \\  
2453943.67176 &  0.73878 & 4060 & 2.16-2.24 & 0.84-0.87 \\
2453871.82770 &  0.77372 & 4060 & 2.32-2.15 & 0.45-0.36 \\
2453626.50659 &  0.81985 & 2700 & 2.23-2.15 & 0.78-0.72 \\
2454003.51981 &  0.87481 & 4060 & 2.15-2.33 & 0.93-0.81 \\
\hline

\end{tabular}
\end{flushleft}
\end{table}
}

%% file: 0749tab7.tex
\begin{table*}
\caption[]{\label{tab:v20_spel}
Spectroscopic elements for V20 determined from the two spectral regions applying
different templates. Both double-lined and single-lined solutions are included.
Epoch and orbital period were fixed at the values listed in Eq.~\ref{eq:v20_eph},
and a circular orbit was assumed.
$K_p$ and $K_s$ are the velocity semi-amplitudes for primary and secondary
component, respectively, $\gamma$ is the system velocity for V20, and
$RV_{3}$ is the mean value of the radial velocity for the 3rd star.
$\sigma$ is the standard error of one observation.
All quantities are given in units of $\kms$.
Spectra regions (R) are: L = 5100-5600\,\AA; U = 6100-6700\,\AA\ with 
$\mathrm{H_{\alpha}}$ blocked out.
Templates (T) used are 
P (Primary):   $T_{\rm{eff}}$ = 5600 K, log($g$) = 4.2, \feh\,$=+0.4$;
S (Secondary): $T_{\rm{eff}}$ = 4800 K, log($g$) = 4.6, \feh\,$=+0.4$;
C (3rd star):  $T_{\rm{eff}}$ = 5000 K, log($g$) = 4.5, \feh\,$=+0.4$.
The solutions marked by $*$ are based on velocities corrected by the results from
the analyses {\bf o}f synthetic spectra; see text for details.
}
\begin{center}
\begin{tabular}{llcccccc} \hline
\hline\noalign{\smallskip}
R       & T        & $K_p$          & $K_s$     & $\sigma_p$& $\sigma_s$ & $\gamma$         & $RV_{3}$\\
\hline\noalign{\smallskip}
L       & P        &$47.04 \pm 0.20$&$61.22 \pm 0.20$&  0.37      & 0.89 &$-46.49 \pm 0.12$ & $-44.43 \pm 0.61$\\
L*      & P*       &$47.01 \pm 0.19$&$61.31 \pm 0.19$&  0.37      & 0.86 &$-46.49 \pm 0.12$ & $-44.43 \pm 0.61$\\
L       & S        &$47.15 \pm 0.20$&$61.13 \pm 0.20$&  0.39      & 0.92 &$-46.46 \pm 0.12$ & $-44.52 \pm 0.57$ \\
L       & C        &$47.13 \pm 0.20$&$61.14 \pm 0.20$&  0.38      & 0.89 &$-46.47 \pm 0.12$ & $-44.52 \pm 0.57$\\
U       & P        &$47.19 \pm 0.11$&$61.24 \pm 0.11$&  0.35      & 0.43 &$-46.56 \pm 0.07$ & $-44.55 \pm 0.73$\\
U*      & P*       &$47.07 \pm 0.11$&$61.18 \pm 0.11$&  0.37      & 0.41 &$-46.50 \pm 0.07$ & $-44.50 \pm 0.73$\\
U       & S        &$47.18 \pm 0.15$&$61.16 \pm 0.15$&  0.36      & 0.62 &$-46.54 \pm 0.09$ & $-44.49 \pm 0.71$\\
U       & C        &$47.19 \pm 0.13$&$61.15 \pm 0.14$&  0.36      & 0.57 &$-46.54 \pm 0.08$ & $-44.49 \pm 0.70$\\
L       & P        &$46.98 \pm 0.08$&                &  0.27      &      &$-46.71 \pm 0.07$ &                  \\
L*      & P*       &$46.96 \pm 0.08$&                &  0.27      &      &$-46.71 \pm 0.07$ &                  \\
L       & S        &$47.08 \pm 0.08$&                &  0.26      &      &$-46.71 \pm 0.07$ &                  \\
L       & C        &$47.06 \pm 0.08$&                &  0.26      &      &$-46.71 \pm 0.07$ &                 \\
U       & P        &$47.17 \pm 0.10$&                &  0.33      &      &$-46.63 \pm 0.09$ &                  \\  
U*      & P*       &$47.04 \pm 0.10$&                &  0.33      &      &$-46.63 \pm 0.09$ &                  \\  
U       & S        &$47.17 \pm 0.11$&                &  0.35      &      &$-46.62 \pm 0.09$ &                  \\
U       & C        &$47.17 \pm 0.10$&                &  0.34      &      &$-46.62 \pm 0.09$ &               \\   
L       & P        &                &$61.16 \pm 0.25$&            & 0.82 &$-46.26 \pm 0.22$ &                  \\ 
L       & S        &                &$61.07 \pm 0.26$&            & 0.84 &$-46.20 \pm 0.22$ &                  \\ 
L       & C        &                &$61.07 \pm 0.25$&            & 0.81 &$-46.22 \pm 0.21$ &                  \\ 
U       & P        &                &$61.22 \pm 0.13$&            & 0.41 &$-46.50 \pm 0.11$ &                   \\ 
U       & S        &                &$61.14 \pm 0.18$&            & 0.56 &$-46.48 \pm 0.16$ &                  \\
U       & C        &                &$61.13 \pm 0.17$&            & 0.54 &$-46.47 \pm 0.14$ &                \\
\noalign{\smallskip}
\hline
\end{tabular}
\end{center}
\end{table*}

%% file: 0749tab8.tex
\begin{table}   
\caption[]{\label{tab:v20_orbit}
Spectroscopic orbital solution for V20.
$T$ is the time of central primary eclipse.
}
\begin{center}    
\begin{tabular}{lr} \hline   
\hline\noalign{\smallskip}    
Parameter            & \multicolumn{1}{c}{Value} \\ 
\noalign{\smallskip}
\hline
\noalign{\smallskip}    
Adjusted quantities:            &   \\ 
$K_p$~(\kms)                     &$ 47.09 \pm 0.09 $  \\   
$K_s$~(\kms)                     &$ 61.16 \pm 0.20 $  \\   
$\gamma$~(\kms)                  &$-46.63 \pm 0.13 $  \\   
\noalign{\smallskip}  
Adopted quantities:             &     \\
$T$~(HJD$-$2\,400\,000)          &  53151.6061  \\
$P$~(days)                      & 14.469918   \\
$e$                             &  0.00          \\ 
\noalign{\smallskip}  
Derived quantities:             &      \\
$M_p \sin^3i~\mathrm{(M_{\sun})}$       & $ 1.074  \pm 0.008  $ \\
$M_s \sin^3i~\mathrm{(M_{\sun}})$       & $ 0.827  \pm 0.004  $ \\
$a \sin i~\mathrm{(R_{\sun})}$          & $30.946  \pm 0.063  $ \\
\noalign{\smallskip}  
Other quantities:              &      \\
$N_{obs}$                      &    15 \\
Time span (days)               &   377 \\
\noalign{\smallskip}  
\hline
\end{tabular}            
\end{center}            
\end{table}                                  

%% file: 0749tab9.tex
\begin{table}   
\caption[]{\label{tab:v20_absdim}
Astrophysical data for V20.
$T_{eff\sun} = 5780$ K, $B.C._{\sun} = -0.08$, and $M_{bol\sun} = 4.74$ 
has been assumed. We have adopted 
$E(B-V) = 0.15  \pm 0.02 $,           
$E(V-I) = 1.3 \times E(B-V)$,
and $A_V = 3.1 \times E(B-V)$.
$v_{sync}$ is the equatorial velocity for synchronous rotation.
}
\begin{flushleft}    
\begin{tabular}{lrr} \hline    
\noalign{\smallskip}    
\hline    
\noalign{\smallskip}    
                     &    Primary       &    Secondary      \\ 
\noalign{\smallskip}    
\hline    
\noalign{\smallskip}    
Absolute dimensions:          &                   &                 
 \\ 
$M/M_{\sun}$                  &$1.074 \pm 0.008$  &$0.827 \pm 0.004$
\\ 
$R/R_{\sun}$                  &$1.399 \pm 0.016$  &$0.768 \pm 0.006$ 
\\ 
$\log g$ (cgs)                & $4.178 \pm 0.010$ & $4.586 \pm 0.008$
\\ 
\\
$v_{sync}$ (\kms)             & $ 4.9 \pm 0.1$    & $ 2.7 \pm 0.1$
\\
 & & \\ 
Photometric data:             &                   &                 
\\
 $V$       &  $17.642 \pm 0.028$  &  $19.867 \pm 0.041$ \\
 $I$       &  $16.729 \pm 0.033$  &  $18.639 \pm 0.046$ \\
 $(V-I)$   &  $ 0.914 \pm 0.008$  &  $ 1.228 \pm 0.044$ \\ 
 $V_0$     &  $17.177 \pm 0.068$  &  $19.400 \pm 0.074$ \\  
 $(V-I)_0$ &  $ 0.719 \pm 0.027$  &  $ 1.031 \pm 0.051$ \\  
          &                 &               \\
 
$T_{\mbox{\scriptsize eff}}\,$ (K)  &  $5665 \pm 100$ &   $4900 \pm 100$ \\
$M_{\mbox{\scriptsize bol}}\,$      &  $4.10  \pm 0.08$  &   $6.03  \pm 0.09$ \\
$\log L/L_{\sun}$ & $0.26 \pm 0.03$ &   $-0.52 \pm 0.04$ \\
$B.C.$            & $-0.10$         &    $-0.35$ \\
$M_V$ &             $ 4.20 \pm 0.08$&   $ 6.38 \pm 0.09$ \\
                 &                  &             \\
$V-M_V$          &$13.44  \pm 0.09 $& $13.48  \pm 0.10 $ \\
$V_0-M_V$        &$12.98  \pm 0.11 $& $13.02  \pm 0.12 $ \\
Distance \, (pc) &$3940   \pm 200  $& $4015   \pm 225  $ \\
\noalign{\smallskip}            
\hline
\end{tabular}            
\end{flushleft}            
\end{table}

%% file: 0749ta10.tex
\begin{table}
\caption[]{\label{tab:v20_age}
Ages for the componets of V20 as determined from $M-R$ isochrones
calculated for different models and chemical compositions.
For each model, the first line represent the available models
closest to the observed \feh. All models include core overshoot;
BaSTI standard and core overshoot models yield nearly identical
results.
Uncertainties due to mass and radius errors are about
0.3 Gyr (primary) and 0.9 Gyr (secondary).}
\begin{center}
\begin{tabular}{llll} \hline
\hline
Model         & [Fe/H]    & Primary   & Secondary \\
\hline
VRSS          & 0.37      & 7.7       & 7.2 \\   
              & 0.23      & 7.8       & 7.0 \\
              & 0.49      & 7.3       & 6.8 \\
$Y^2$         & 0.40      & 8.2       & 6.2 \\   
              & 0.30      & 8.3       & 6.2 \\
              & 0.50      & 7.8       & 6.2 \\
BaSTI         & 0.395     & 9.0       & 8.6 \\   
              & 0.254     & 8.5       & 7.6 \\
\hline
\end{tabular}
\end{center}
\end{table}

%% file: 0749MAIN.bbl
\begin{thebibliography}{}
\bibitem[2007]{attm07}                                  
Anthony--Twarog, B.J., Twarog, B.A., Mayer, L., 2007, 
\aj, 133, 1585 (ATTM07)

\bibitem[2005]{bedin05} Bedin, L.~R., Salaris,         
M., Piotto, G., King, I.~R., Anderson, J., Cassisi, S., 
\& Momany, Y.\ 2005, \apjl, 624, L45 

\bibitem[2008]{bedin08-1}                              
Bedin, L.~R., King, I.~R., Anderson, J., Piotto, G., Salaris, M., 
Cassisi, S., \& Serenelli, A.\ 2008, \apj, 678, 1279 

\bibitem[2003]{hb03}                                    
Bruntt, H., Grundahl. F., Tingley, B. et al. 2003,
\aap, 410, 323
\bibitem[2006]{psicen}                                 
Bruntt, H., Southworth, J., Torres, G. et al. 2006, 
\aap, 456, 651
\bibitem[2005]{carney05}                               
Carney, B.~W., Lee, J.-W., \& Dodson, B. 2005,
\aj, 129, 656
\bibitem[2006]{carraro06}                               
Carraro, G., Villanova, S., Demarque, P. et al.
2006, \apj, 643, 1151
\bibitem[2007]{carretta07}                              
Carretta, E., Bragaglia, A., \& Gratton, R.~G.
2007, \aap, 473, 129
\bibitem[2006]{cas06}                                   
Casagrande, L., Portinari, L., \& Flynn, C.
2006, \mnras, 373, 13
\bibitem[2000]{c00}                                     
Claret, A., 2000, \aap, 363, 1081
\bibitem[2004]{clausen04}                               
Clausen, J.~V. 2004, New Ast. Rev., 48, 679
\bibitem[1999]{granada99}
Clausen,  J.~V., Baraffe,  I., Claret,  A., \&
VandenBerg,  D.~B. 1999,
in Theory and Tests of Convection in Stellar
Structure, ed. A. Gim\'enez, E. F. Guinan, \& B. Montesinos,
ASP Conference Series, 173, 265
\bibitem[2008]{avw08}                                   
Clausen, J.~V., Torres, G., Bruntt, H., et al. 2008
\aap, 487, 1095


\bibitem[2007]{demarchi07}   
de Marchi, F., et al.\ 2007, \aap, 471, 515  
\bibitem[2000]{dekker00}  
Dekker, H., D'Odorico, S., Kaufer, A., Delabre, B., \& Kotzlowski, 
H.\ 2000, \procspie, 4008, 534 

\bibitem[2004]{yale04}
Demarque, P., Woo, J.-H., Kim, Y.-C., \& Yi, S.~K.   
2004, \apjs, 155, 667
\bibitem[1981]{e81}                                    
Etzel P.~B. 1981, in Photometric and Spectroscopic 
Binary Systems. eds. E.B. Carling and Z. Kopal, 
p. 111, Reidel Publishing Company, Dordrecht
\bibitem[2004]{e04}                                     
Etzel, P.~B. 2004, SBOP: Spectroscopic Binary
Orbit Program (San Diego State University)
\bibitem[1996]{flower96}                               
Flower, P.~J. 1996, \apj, 469, 355
\bibitem[1993]{gn93}                                   
Grevesse, N., \& Noels, A. 1993, Phys. Scr. T47, 133
\bibitem[1996]{gns96}                                  
Grevesse, N., Noels, A., \& Sauval, A.~J. 1996,
in Cosmic Abundances, eds. S.S. Holt \& G. Sonneborn
(San Francisco: ASP), 117
\bibitem[1993]{vh93}                                    
Van Hamme, W.  1993, \aj, 106, 2096
\bibitem[2002]{heiter02}                               
Heiter, U., Kupka, F., van't Menneret, C. et al.
2002, \aap 392, 619


\bibitem[2006]{kaluzny06}
Kaluzny, J., Pych, W., Rucinski, S.~M., \& Thompson, I.~B, 2006, AcA, 56, 237

\bibitem[2002]{yale2}                                  
Kim, Y.-C., Demarque, P., Yi, S.~K., \& Alexander, D.~R.
2002, \apjs, 143, 499
\bibitem[1965]{kinman65}                                
Kinman, T.~D. 1965, \apj, 142, 655
\bibitem[2006]{korn06} Korn, A.~J., Grundahl, F., 
Richard, O., Barklem, P.~S., Mashonkina, L., Collet, R., Piskunov, N., 
\& Gustafsson, B.\ 2006, \nat, 442, 657


\bibitem[1956]{kvw56}                                  
Kwee, K.~K., \& van Woerden, H. 1956, BAN, 12, 327

\bibitem[2008]{lacy08} 
Lacy, C.~H.S., Torres, G. \& Claret, A.\ 2008, \aj, 135, 1757


\bibitem[1998]{landsman98}   
Landsman, W., Bohlin, R.~C., Neff, S.~G., O'Connell, R.~W., Roberts, 
M.~S., Smith, A.~M., \& Stecher, T.~P.\ 1998, \aj, 116, 789 

\bibitem[1994]{liebert94}    
Liebert, J., Saffer, R.~A., \& Green, E.~M.\ 1994, \aj, 107, 1408 

\bibitem[1973]{m73}                                    
Martynov D.~Ya. 1973, in Eclipsing Variable Stars, 
ed. V.P. Tsesevich, Israel Program for Scientific 
Translation, Jerusalem
\bibitem[2002]{m02}                                    
Mochejska, B.~J., Stanek, K.~Z., Sasselov, D.~D., \&
Szentgyorgti, A.~H., 2002, \aj,  123, 3460

\bibitem[2005]{mochejska05} 
Mochejska, B.~J., Stanek, K~.Z., Sasselov, D.~D.
et al.\ 2005, \aj, 129, 2856 

\bibitem[2007]{montalto07}    
Montalto, M., Piotto, G., Desidera, s. et al.\ 2007, 
\aap, 470, 1137 

\bibitem[1972]{nd72}
Nelson B., \& Davis W. 1972, \apj, 174, 617            

\bibitem[2006]{origlia06}                              
Origlia, L., Valenti, E., Rich, R.~M., \& Ferraro, F.~R.
2006, \apj, 646, 499

\bibitem[2008]{pasquini08}                                
Pasquini, L., Biazzo, K., Bonifacio, P., Randich, S., \& Bedin, L.
\aap, Accepted, {arXiv: 0807.0092} 

\bibitem[2004]{basti04}                                
Pietrinferni, A., Cassisi, S., \& Salaris, M.
2004, Mem. S.A.It. 75, 170

\bibitem[1980]{dmp80}                                  
Popper, D.~M. 1980, \araa, 18, 115

\bibitem[1997]{dmp97}                                  
Popper, D.~M. 1997, AJ, 114, 1195
\bibitem[1981]{pe81}
Popper D.~M., \& Etzel P.~B. 1981, \aj, 86, 102             
\bibitem[1992]{press92}
Press, W.~H., Flannery, B.~P., Teukolsky, S.~A.,
\& Vetterling, W.~T. 1992, Numerical Recepies in Fortran 77,
2nd ed. Cambridge Univ. Press, Cambridge
\bibitem[2005]{rm05}                                   
Ram\'\i rez, I., \& Mel\'endez, J. 2005, \aj, 626, 465

\bibitem[1996]{rucinski96}   
Rucinski, S.~M., Kaluzny, J., \& Hilditch, R.~W.\ 1996, \mnras, 282, 70

\bibitem[1999]{r99}                                     
Rucinski, S.~M. 1999, in 
Hearnshaw, J.~B. \& Scarfe, C.~D. (eds) IAU Coll. 170
Precise Stellar Radial Velocities, ASP Conference
Series 185, 82
\bibitem[2002]{r02}                                     
Rucinski, S.~M. 2002, \aj, 124, 1746
\bibitem[2004]{r04}                                     
Rucinski, S.~M. 2004, in
Maeder, A. \& Eenens, P. (eds) IAU Symp. 215
Stellar Rotation, ASP Conference Series, p.17
\bibitem[1998]{sch98}                                   
Schlegel, D.~J., Finkbeiner, D.~P., \& Davis, M. 1998,
\apj\, 500, 525

\bibitem[2004]{southworth04}                          
Southworth, J., Maxted, P.~F.~L., \& Smalley, B. 2004,
  \mnras, 349, 547 

\bibitem[2004a]{sms04}                                  
Southworth, J., Maxted, P.~F.~L., \& Smalley, B. 2004a,
  \mnras, 351, 1277 

\bibitem[2004b]{szm04}                                  
Southworth, J., Zucker, S., Maxted, P.~F.~L., 
\& {Smalley}, B.  2004b, \mnras, 355, 986 

\bibitem[2007]{betaaur}                                 
Southworth, J., Bruntt, H., \& Buzasi, D.~L. 2007,
\aap, 467, 1215
\bibitem[1987]{stetson87}                               
Stetson, P.~B. 1987, \pasp, 99, 191
\bibitem[1990]{stetson90}                               
Stetson, P.~B. 1990, \pasp, 102, 932
\bibitem[1994]{stetson94}                               
Stetson, P.~B. 1994, \pasp, 106, 250 
\bibitem[2003]{stetson03}                               
Stetson, P.~B., Bruntt, H., \& Grundahl, F. 2003, 
\pasp, 115, 413 (SBG03)
\bibitem[2006]{wt06}                                   
Torres, G., Lacy, C.H., Marschall, L.~A.,
Sheets, ,H.~A, \& Mader, J.~A. 2006, \aj, 640, 1018

\bibitem[1995]{tripicco95}               
Tripicco, M.~J., Bell,  R.~A., Dorman, B., \& Hufnagel, B.\ 1995, \aj, 109, 1697 
\bibitem[2000]{vdb00}                                   
VandenBerg, D.~A., Swenson, F.~J., Rogers, F.~J.,
Iglesias, C.~A., \& Alexander, D.~R. 2000, \apj, 532, 430
\bibitem[1996]{vp96}                                   
Valenti, J., \& Piskunov, N. 1996, \aaps, 118, 595
\bibitem[2004]{vdb04} VandenBerg, D.~A., \& Stetson, P.~B.\ 2004, \pasp, 116, 997
\bibitem[2006]{vr06}                                    
VandenBerg, D.~A., Bergbusch, P.~A., \& Dowler, P.~D. 2006
\apjs, 162, 375

\end{thebibliography}
